\documentclass[referee]{raa}            
\usepackage{tikz}
\usepackage{multirow}      
\usepackage{array}         
\usepackage{caption}       
\usepackage{booktabs}      
\usepackage{graphicx,times}             
\usepackage{natbib}                     
\usepackage{amssymb,amsmath}              
\bibpunct{(}{)}{;}{a}{}{,}                
\usepackage{booktabs}  
\usepackage{arydshln} 

\usepackage[pagebackref=true]{hyperref}  
\usepackage{multirow}

\begin{document}

  \title{Attitude Estimation of Falcon 9 Rocket Body Based on Photometric Characteristics
}

   \volnopage{Vol.0 (20xx) No.0, 000--000}      
   \setcounter{page}{1}          

   \author{Hui Li  
      \inst{1,2}
   \and Rong-Wang Li
      \inst{1,3}
   \and Ru-Feng Tang
      \inst{1}
   \and Yu-Qiang Li
      \inst{1,3}
   }
   \institute{Yunnan Observatories, Chinese Academy of Sciences, Kunming 650216, China
   \and
   University of Chinese Academy of Sciences, Beijing 100049, China; {\it lirw@ynao.ac.cn}\\
   \and
   Key Laboratory of Space Object and Debris Observation,Chinese Academy of Sciences,Nanjing 210023, China
\vs\no
   }

\abstract{ 
The proliferation of space debris, driven by increasingly frequent space activities, poses a severe risk to operational satellites and future missions. Consequently, the surveillance and characterization of space objects have become critical. Current methods for attitude estimation of space objects often rely on target-specific models, such as those with a closed set of facets, and typically employ nonlinear filtering or machine learning techniques. However, these approaches often lack general applicability. This study proposes a more general approach to attitude estimation based on photometric light curves. We define a loss function that quantifies the discrepancy between the measured light curve and a theoretical model parameterized by the attitude parameters. By leveraging automatic differentiation to compute the gradients of this function, we use a gradient descent algorithm to achieve rapid and accurate attitude estimation. The proposed method provides a versatile and efficient framework for attitude estimation, overcoming the target-specificity limitations of existing techniques, with the parameters estimation time reduced to 15 seconds and the mean square error lowered to three decimal places. Validation on the third-party database demonstrates that the method exhibits strong generalization capability.
}

   \authorrunning{H. Li, R.-W. Li, R.-F. Tang, \& Y.-Q. Li }            
   \titlerunning{Attitude Estimation of Rocket Body}  

   \maketitle

\section{Introduction}        
\label{sect:intro}
The growing population of space debris represents a critical threat to the safety and sustainability of orbital operations. Among this debris, spent rocket upper stages are of particular concern. Their combination of large mass, significant size, and uncontrolled rotational state makes them a primary source of future debris through potential collisions or fragmentation. Precisely determining the attitude of these objects is crucial for mitigating these risks. In the context of orbit prediction, knowledge of an object's attitude is essential for accurately modeling non-gravitational forces, such as atmospheric drag and solar radiation pressure, leading to more reliable trajectory forecasts. Similarly, active debris removal (ADR) missions—including concepts like laser ablation and robotic capture—depend on precise attitude data to ensure safe and successful target engagement. Given these critical applications, the attitude determination of large, non-cooperative objects like rocket bodies has emerged as a key research priority within the field of Space Situational Awareness (SSA).

Foundational work in this area began as early as 1963, when \cite{Giese+1963} investigated the attitude of cylindrical satellites by modeling both specular and diffuse reflection. Williams later developed a theoretical framework for determining the orientation of spinning cylindrical satellites using optical data (\citealt{Williams+1979a}, \citealt{Williams+1979b}).
Building on these principles, subsequent work has applied specific estimation algorithms to observational data. For instance, \cite{Wetterer+Jah+2009} used the Unscented Kalman Filter (UKF) to estimate the attitude of the Centaur rocket body, which has been assigned ID 12069 by North American Aerospace Defense Command (NORAD), from its light curves. In a similar study, \cite{Yanagisawa+Kurosaki+2012} performed shape and attitude estimation for a low-orbit rocket body (NORAD ID 20625) using light curve analysis. Assuming a diffuse reflection model for a cylindrical shape, \cite{Santoni+etal+2013} determined the spin period and axis orientation for two targets—the CZ-3B and SL-12 rocket bodies—based on ground-based optical observations.
The field has also seen a trend toward integrating multiple data sources and developing more sophisticated physical and statistical models. For instance, \cite{Zhao+etal+2020} combined laser ranging data with photometric light curves from the Graz station to precisely determine the spin period and axis orientation of the CZ-3B rocket body (NORAD ID 38253). 
Focusing on modeling, \cite{Abercrombie+etal+2021} developed a physics-based theoretical model to generate light curves for tumbling objects, which they validated using observations of the retired GOES 8 satellite. Statistical approaches have also become prominent. \cite{Campbell+etal+2022} employed a Bayesian framework with Markov Chain Monte Carlo (MCMC) sampling to estimate the spin state and reflection characteristics of the object 2020 SO. In a comparative study, \cite{Clark+etal+2022} evaluated two Bidirectional Reflectance Distribution Function (BRDF) models—a defined facet model and the Phong model. They then utilized the Particle Swarm Optimization (PSO) algorithm to accurately estimate the target's shape, attitude, and optical properties.

Researchers in China have also made significant contributions to the field. \cite{Wang+etal+2017} from the Equipment Academy surveyed recent advances in space object characteristic inversion methods, categorizing the techniques into two primary approaches: those based on light curve features and those relying on geometric facet models. At the Purple Mountain Observatory, \cite{Lin+Zhao+2018} re-estimated the rotational state of the Envisat satellite, incorporating the effects of gravity-gradient-induced precession. \cite{Tang+etal+2018} from Yunnan
Observatories(YNAO) of the Chinese Academy of Sciences (CAS) used a superconducting nanowire single-photon detector (SNSPD) to simultaneously collect laser ranging and photometric data for targets like Topex and CZ-4 DEB, which they used to analyze their rotation periods. Using a different data type, \cite{Liu+etal+2019} from the National Astronomical Observatories of CAS estimated the tumbling attitude of Soyuz rocket bodies using diffuse reflection laser ranging data from the Wettzell station. More recently, returning to machine learning, \cite{Lu+Zhao+2020} (Purple Mountain Observatory) investigated methods for classifying the basic shapes of space debris from light curves, validating their algorithms with both simulated and measured data. Along this data-driven direction, \cite{Zhao+Sun+Yu+2024} developed an end-to-end deep neural network closed-loop framework for optical resident space object detection directly from raw CCD data, demonstrating that high detection performance and improved efficiency can be achieved through data fusion without hardware modification. Such developments in raw-data processing and reduction also benefit from standardized astronomical software tools; in this context, \cite{Ji+etal+2023} presented PyMsOfa, a Python implementation of the IAU SOFA service for astronomical calculations.

In a relevant study, \cite{Su+etal+2021} (YNAO) estimated the attitude of a Falcon 9 upper stage using photometric data from a 1.2m telescope. They first constructed a light curve by normalizing the apparent magnitude for distance and then determined the synodic period using the Phase Dispersion Minimization method. Next, they developed a facet-based photometric model to generate theoretical light curves. This model incorporated the observation geometry (sun, target, and station) and the target's rotational state, including its spin axis orientation and initial phase.
To find the best-fit parameters, they performed an exhaustive grid search to minimize the Mean Squared Error (MSE) between the theoretical and observed curves. While effective, this grid search approach proved to be computationally expensive, requiring approximately one day to yield only a limited set of optimal parameters. This computational bottleneck makes the method unsuitable for rapid attitude estimation. This paper directly addresses this limitation by replacing the inefficient grid search with a gradient-based optimization approach enabled by Automatic Differentiation (AD).

This study proposes an Attitude Estimation method based on Automatic Differentiation (AEAD) to overcome the computational limitations of previous grid-search techniques. Our approach begins by constructing a loss function that quantifies the difference between observed light curves and those generated by a theoretical photometric model. Because an analytical expression for the gradient of this complex function is intractable, we employ AD to compute it efficiently. This enables the use of a gradient descent algorithm to rapidly and accurately optimize for the target's attitude parameters, including its rotation axis orientation.
We validate the performance of the AEAD method on two key datasets: first, on simulated light curves with known ground-truth parameters, and second, on real observational data for the same target from a third-party database. Furthermore, our analysis of the optimization's solution space reveals a approximate linear relationship between the estimated rotation axis orientation ($\lambda$) and the initial phase ($\phi_0$), providing new insight into the target's rotational dynamics.


\section{DATA}
\label{sect:data}
\subsection{Target information}
\begin{figure}
\centering
    \includegraphics[width=10cm, keepaspectratio]{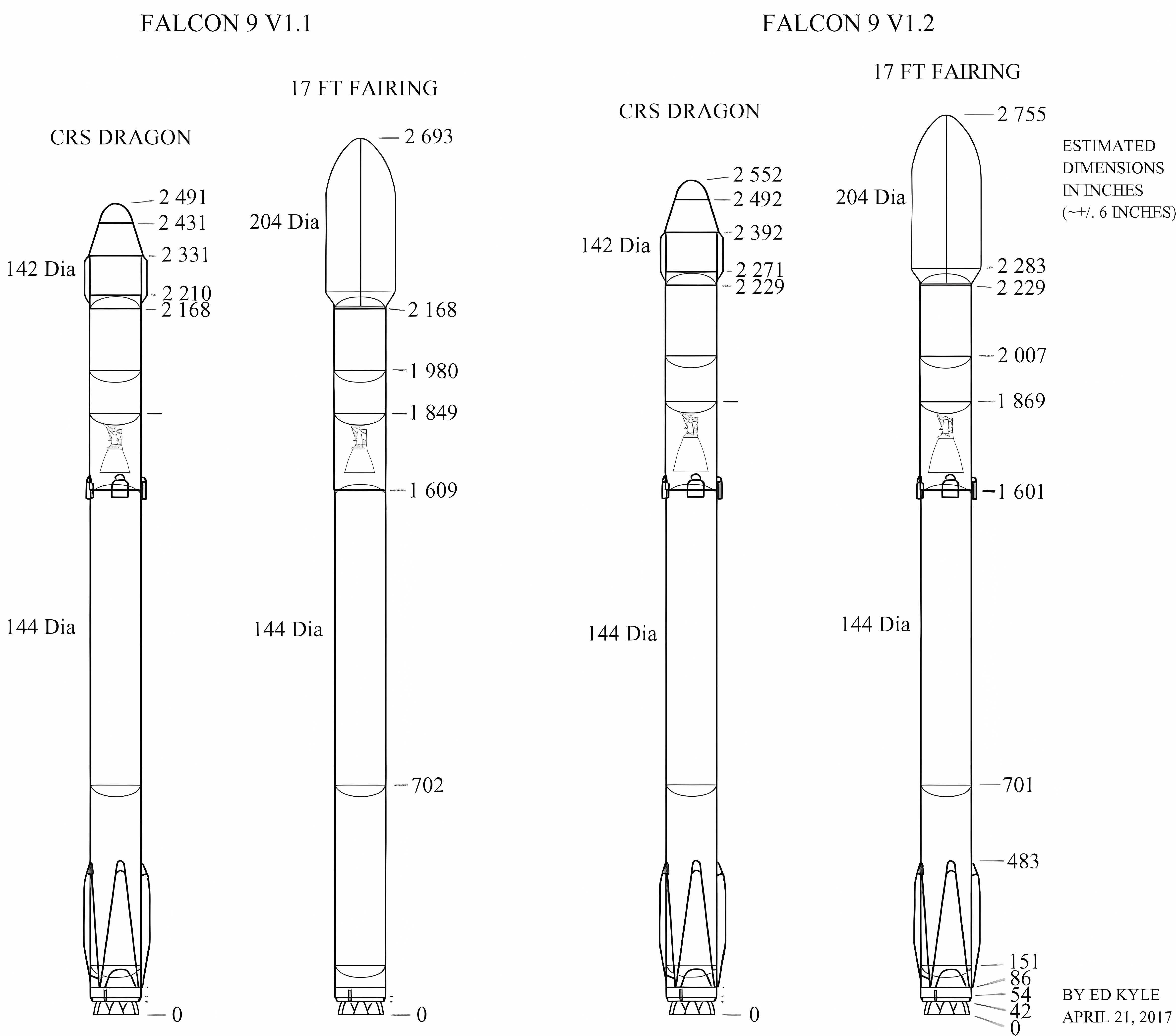}
    \caption{The object explored in this study is depicted on the far right: the upper stage of the Falcon-9 V1.2 rocket. It measures approximately 9.44 meters in length after removing the bottom engines and has a bottom diameter of about 3.66 meters.}
    \label{fig:1}
\end{figure}

The target for this study is the upper stage of a Falcon 9 v1.2 rocket (NORAD ID 42968), hereafter referred to as F9 RB. This object was selected as it represents a common and significant class of space debris. Due to its large dimensions, the F9 RB is an excellent candidate for photometric analysis, making it a suitable target for validating our attitude estimation methodology. The physical dimensions of the stage, based on publicly available data from SpaceX, are illustrated in Figure~\ref{fig:1} \footnote{https://sma.nasa.gov/LaunchVehicle/assets/spacex-falcon-9-v1.2-data-sheet.pdf}.

\subsection{Simulated light curves with known target rotation axis orientation}
\begin{figure}
    \centering
    \includegraphics[width=10cm, keepaspectratio]{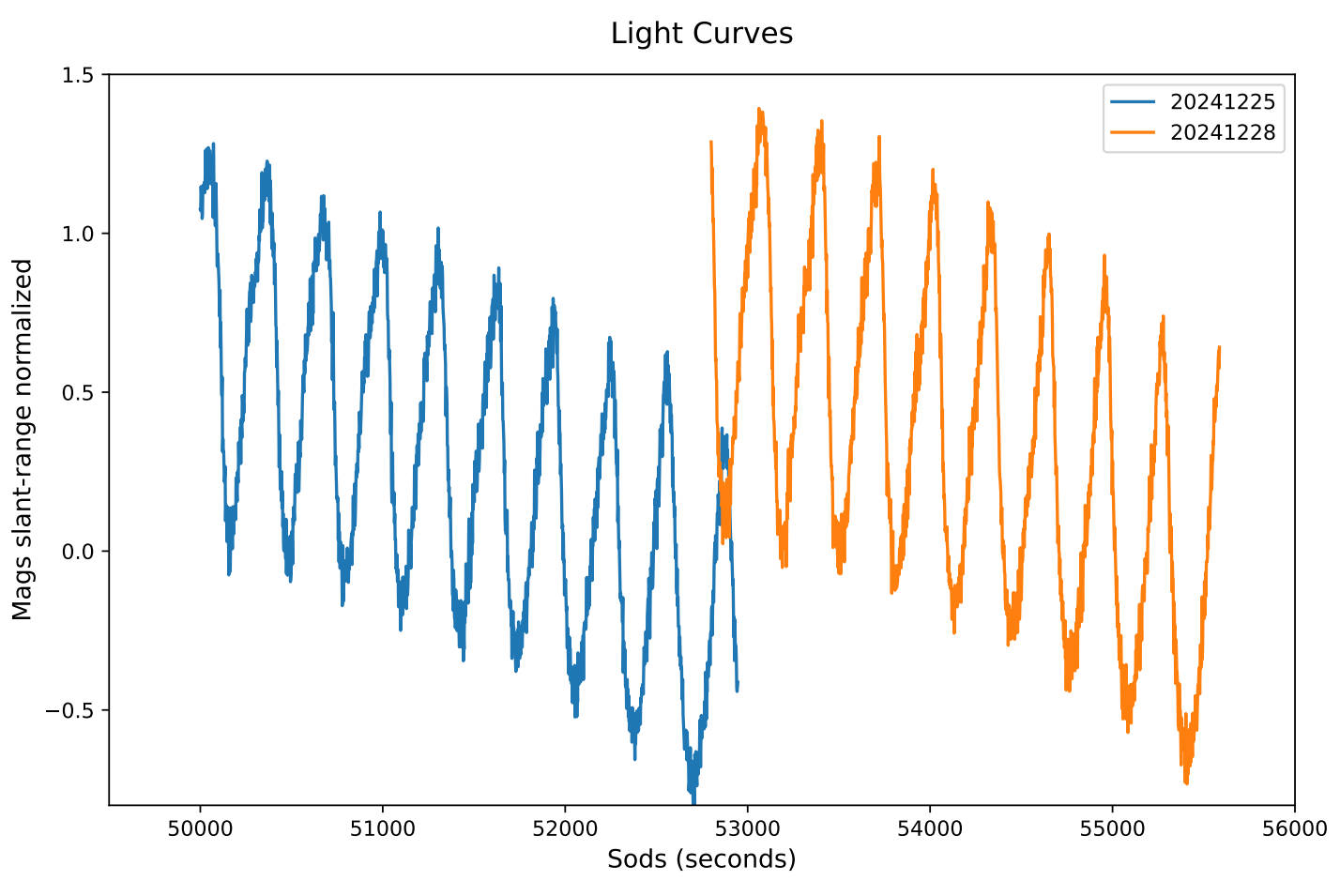}
    \caption{Simulated light curve with rotation axis orientation at $\lambda = 200^\circ$ and $\theta = 90^\circ$. The horizontal axis represents seconds of the day, while the vertical axis shows the slant-range normalized magnitudes.}
    \label{fig:3}
\end{figure}

To validate the AEAD method, we generated simulated light curves  using the photometric model detailed in Section 3.2. This model  simplifies the target (NORAD ID 42968) into a cylinder and computes its  theoretical magnitude by considering the illumination geometry between  the Sun, the target, and the observer. The ground-truth spin axis  orientation for the simulation was set to a right ascension of 200° and a declination of 90°. To mimic realistic observation conditions, Gaussian noise was added to the synthetic data. The scenario was simulated for a ground station at 25.0299°N latitude, 102.7974°E longitude, and  1991.83 m altitude, considering the target's visibility. Two observation windows were simulated: from 13:53:20 to 14:42:29 UTC on December 25,  2024, and from 14:40:00 to 15:26:27 UTC on December 28, 2024. The  resulting light curves used for this validation are presented in  Figure~\ref{fig:3}.

\subsection{Target's light curves measured by Yunnan Observatories}
\begin{figure}
	\includegraphics[width=\columnwidth]{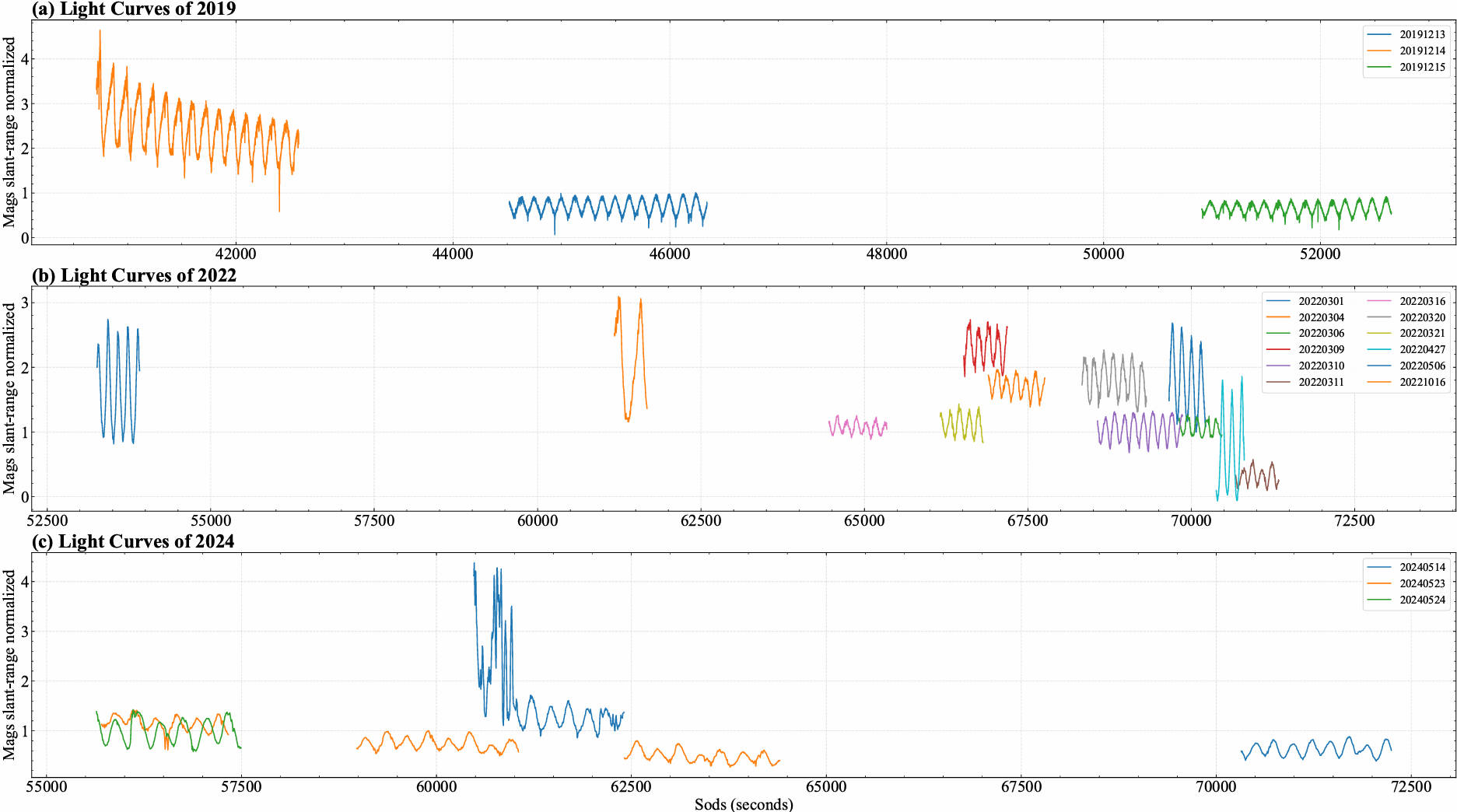}
    \caption{Multi-epoch observed light curves of target 42968, with different colors corresponding to distinct observation dates.}
    \label{fig:2}
\end{figure}

The primary observational data used in this study were acquired with  the 1.2-meter telescope at YNAO during campaigns in 2019, 2022, and 2024. The telescope is equipped  with an Andor DU888 1024x1024 pixel CCD, providing a 36' x 36' field of  view. While also capable of high-precision laser ranging, the instrument is primarily dedicated to the routine photometric monitoring of space  debris, for which YNAO maintains an extensive database. The full set of  light curves for our target, F9 RB, is presented in  Figure~\ref{fig:2}. The figure presents measured light-curve data obtained from a total of 18 separate transits across three different years. Each observation lasts approximately 2000 seconds, and most light curves contain about 3–8 rotation periods. For this study, the measured light curves must have sufficiently long and continuous observation durations to ensure an adequate number of photometric data points. Therefore, observations that meet this requirement—such as the light curve measured on 2019-12-13—are used in the present work.

Observation planning for each session began  with orbit prediction using the target's Two-Line Element (TLE) data to  determine its transit times (rise and set) for the YNAO station  coordinates. An elevation mask was applied to define the effective  observation window, thereby avoiding interference from terrain and high  atmospheric extinction. The raw CCD images were processed through a standard photometric data reduction pipeline to generate the final  light curves. This pipeline consisted of these main steps: The first step was to apply background correction in order to remove instrumental noise and dark current. This was followed by flat-field processing, which compensated for the non-uniform response of individual CCD pixels. After these preprocessing procedures, slant-range normalization was performed to eliminate the effects of varying observational geometry. The resulting normalized time-series light curve is the core data product used for the analysis presented in this paper.

\subsection{Light curves measured by MMT}
\begin{figure}
    \centering
    \includegraphics[width=10cm, keepaspectratio]{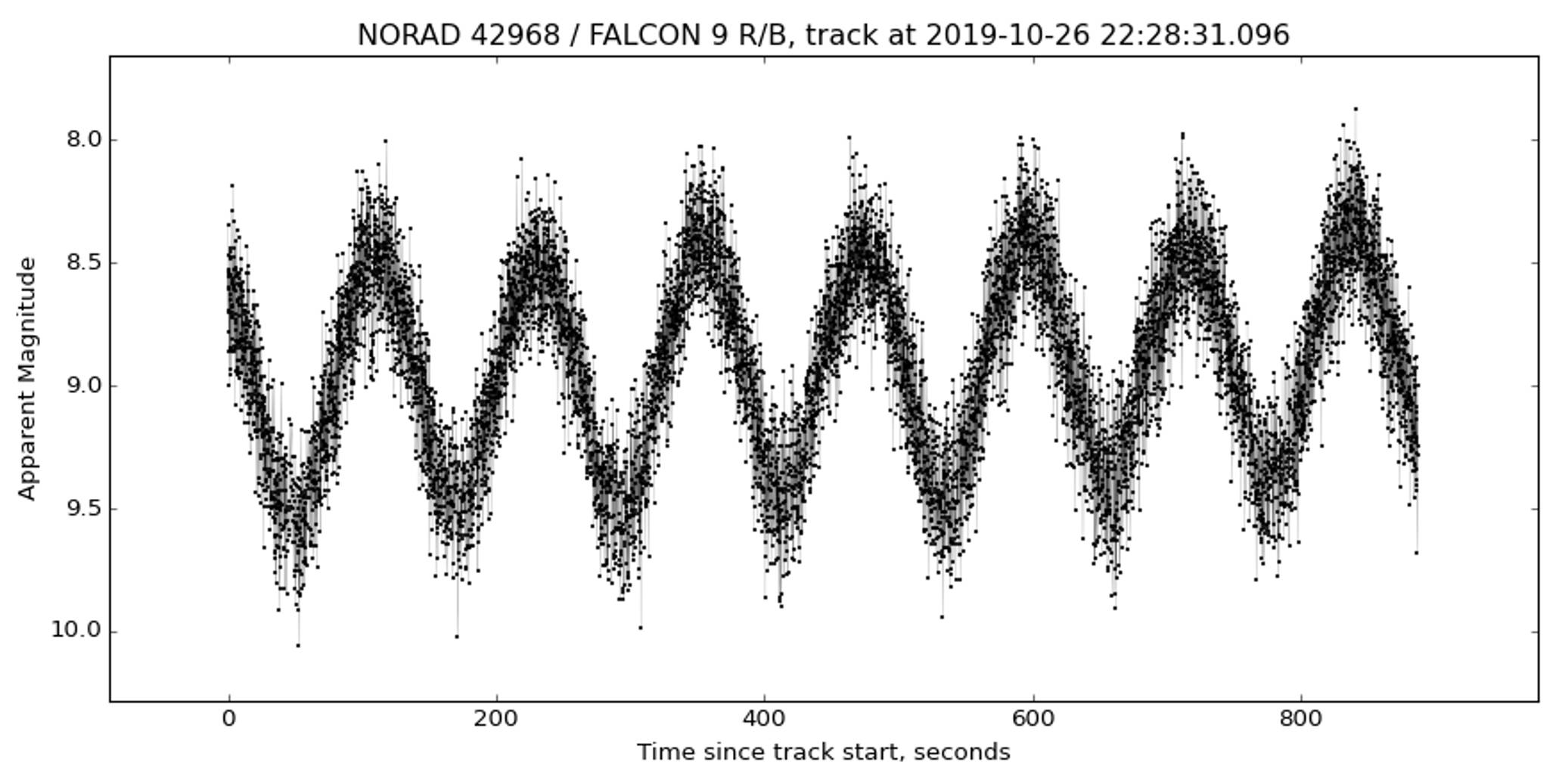}
    \caption{
MMT-observed photometric measurements of the target (2019-10-26), plotted as
 apparent magnitude versus sods.
    }
    \label{fig:4}
\end{figure}

\begin{figure}
    \centering
    \includegraphics[width=10cm, keepaspectratio]{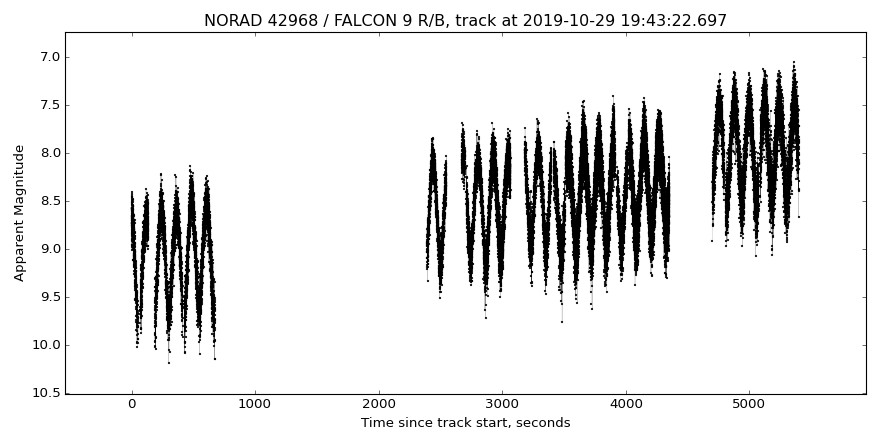}
    \caption{
    MMT-observed photometric measurements of the target (2019-10-29).
    }
    \label{fig:5}
\end{figure}

To demonstrate the generalizability of the AEAD method, we applied it to an independent dataset of measured light curves from the Mini-MegaTORTORA (MMT) system \footnote{http://mmt.favor2.info/satellites}. The MMT system (\citealt{Karpov+etal+2016}) is a novel multi-purpose wide-field monitoring instrument built for and owned by the Kazan Federal University, presently operated under an agreement between Kazan Federal University and Special Astrophysical Observatory, Russia. The novelty of the MMT is its ability to re-configure itself from a wide-field to narrower-field instrument, which may open new ways of studying the sky, as it may, in principle, autonomously perform thorough study of objects it discovers - to simultaneously acquire threecolor photometry and polarimetry of them. The light curves obtained from this system are presented in Figures~\ref{fig:4} and~\ref{fig:5}.

\section{Method}
\label{sect:classification method}
\subsection{Hypotheses related to the target photometry model}
\begin{figure}
\centering
	\includegraphics[width=8cm, keepaspectratio]{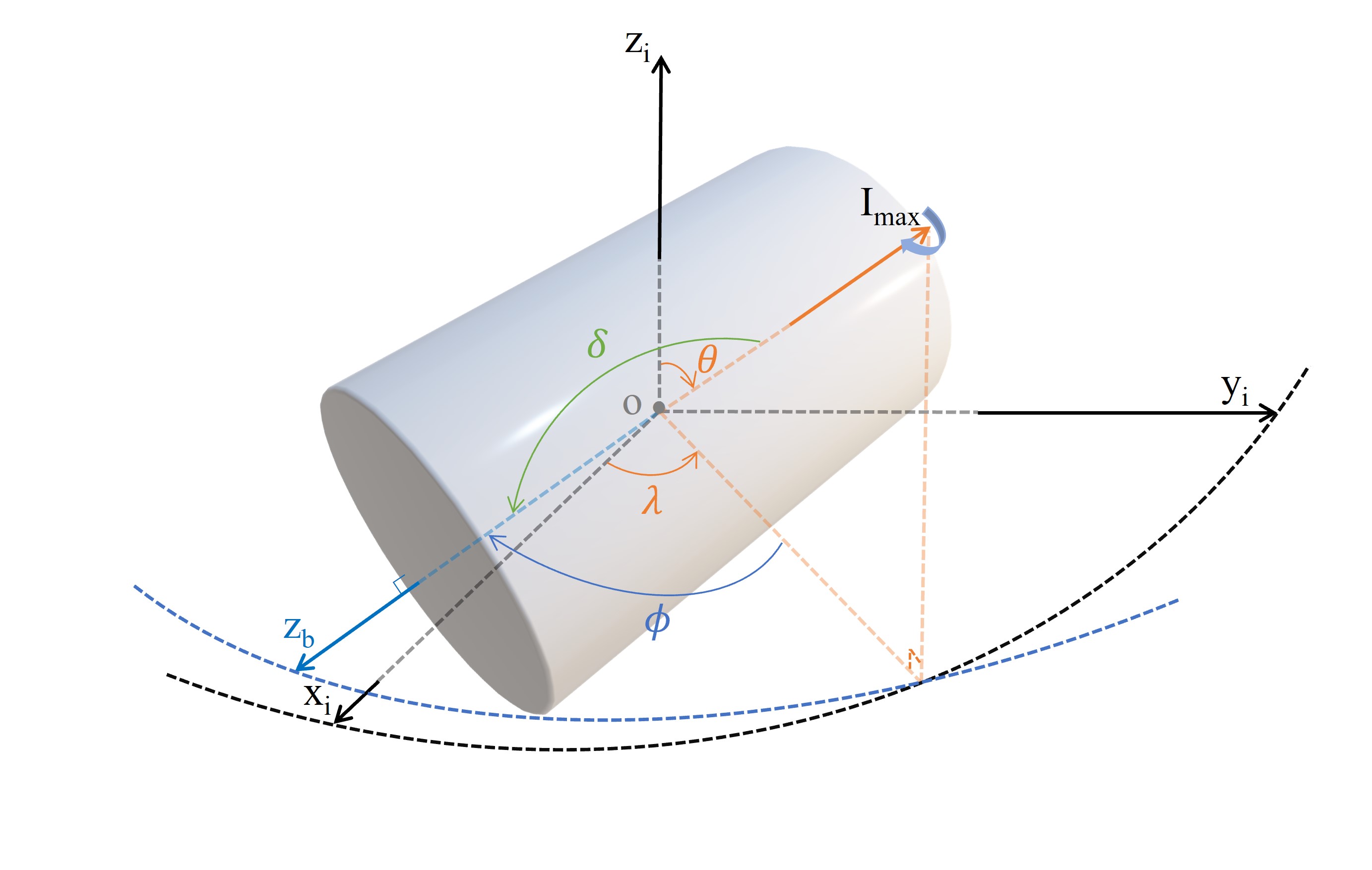}
    \caption{Cylindrical model of the F9 RB. $O-x_i y_i z_i$ is the inertial coordinate system, and $z_b$ axis shown in blue represents the z-axis of the body coordinate system. $I_{max}$ represents the axis of maximum moment of inertia.}
    \label{fig:6}
\end{figure}

The rotational dynamics of a space object are governed by a complex interplay of its physical properties (shape, mass distribution, material) and external environmental torques (e.g., gravity gradient, solar radiation pressure, aerodynamic drag). To create a tractable analytical model for attitude estimation, we adopt the following simplifying assumptions:
\begin{enumerate}
    \item The rocket body is treated as a rigid cylinder.
    \item The object is assumed to be in a stable spin state, rotating purely about its principal axis of maximum moment of inertia (sometimes known as the tennis racket theorem; see Figure~\ref{fig:6}).
    \item The surface is modeled as a Lambertian reflector with a diffuse reflection coefficient, $\rho$, that is constant across the visible light spectrum.
    \item The orientation of the spin axis is assumed to be fixed in inertial space over the observation period of several days.
\end{enumerate}

\subsection{Establishment of the target photometry model}
\label{sec:3.2}
The apparent magnitude of a space object observed from a ground station is determined by its intrinsic properties (shape, surface material, attitude) and the observation geometry, which encompasses the relative positions of the Sun, the target, and the observer (\citealt{Su+etal+2021}).

Our photometric model begins by establishing an attitude rotation matrix to define the transformation from the target's body-fixed coordinate system to the J2000 inertial frame. We model the cylindrical target as three primary surface components: its two circular ends and the lateral surface. Based on the Sun-target-observer geometry, we then calculate the illuminated and visible area of each component from the observer's perspective. The total theoretical magnitude is computed by summing the brightness contributions from all visible components, which constitutes our complete photometric model.

\subsubsection{Rotation matrix and geometry}
The rotation matrix translating the body coordinate system to the inertial system includes the parameters of the rotation axis direction and the  initial phase of rotation. Using the rotation matrix, the normal  direction of each voxel in the body coordinate system can be expressed  in the inertial frame at any given time. According to Fig.~\ref{fig:6}, the  rotation matrix can be represented by the following equation:
\begin{equation}\label{旋转矩阵}
    R_{b2i} = R_z(-\lambda)R_y(-\theta)R_z(-(\pi-\phi(t))R_y(\delta)
\end{equation}
where $\phi(t)$ satisfies:
\begin{equation}
    \phi(t)=\phi_0+\frac{2\pi}{T}(t-t_0)
\end{equation}
where $\lambda$ and $\theta$ are the direction of the rotation axis in the inertial frame. $\phi, \phi_0$ are phase and initial phase of rotation. $\lambda, \theta, \phi_{0}$ are the unknowns to be estimated in this study. $\delta$ is the angle between the axis of maximum moment of inertia $I_{max}$ and the reference positive direction of one underside, set to $90^\circ$. $T$ is the apparent rotation period. The reference positive direction of the underside in the body coordinate system is $(0, 0, 1)$. In Fig.~\ref{fig:6}, the blue dashed curve represents the trajectory of the $z_b$-end when the target rotates about $I_{max}$, while the black dashed curve lies in the $x_ioy_i$ plane.

The image of the target observed by the ground station is actually the reflected light from the rocket's upper stage, which is related to the geometry of the Sun and the station. The phase angle refers to the angle between the Sun, target, and station, and it is calculated as follows:
\begin{equation}
    \Phi = arccos(\frac{\vec{u}_{obs} \cdot \vec{u}_{sun}}{|\vec{u}_{obs}| |\vec{u}_{sun}|})
\end{equation}
The vector $\vec{u}_{obs}$ represents the direction from the target to the station, $\vec{u}_{sun}$ represents the direction from the target to the Sun. The angle between $\vec{u}_{obs}$ and $\vec{u}_{sun}$ is the phase angle $\Phi$.

\subsubsection{Calculating theoretic light curves considering the illumination at the observatory and equivalent areas of the target}
It is also necessary to analyze both the irradiance from the sun to the target and the irradiance from the target to the station. According to Lambert's cosine law (\citealt{Wang+etal+2008}), the irradiance $dE_s$ at the station from the scattering of sunlight by the plane element $dS$ of the target can be calculated: 
\begin{equation}
    dE_s = \frac{E_{0,V}}{\pi r^2}\rho\cos{\theta_i}\cos{\theta_o} dS
\end{equation}
Where $E_{0,V}$ denotes the irradiance of the sun on the target in the $V$ band, $r$ represents the slant distance from the target to the station, $\rho$ is the diffuse reflection coefficient, $\theta_i$ is the angle between the direction of sunlight and the voxel normal, and $\theta_o$ is the angle between the direction of the station and the plane element normal.
To clarify our approach, we introduce the concept of equivalent area. The formula for calculating it is as follows:
\begin{equation}\label{equ: 4}
    A_e = \int \cos{\theta_i}\cos{\theta_o} dS
\end{equation}

Therefore, the ground irradiance generated by the scattering of sunlight on the surface of the target is obtained by summing the irradiance of all the individual voxels.

\begin{equation}\label{equ: 5}
    E_s = \frac{E_{0,V}}{\pi r^2}\sum_{i=1}^{N}\rho_iA_{e,i}
\end{equation}
In equation (\ref{equ: 5}), $A_{e,i}$ represents the equivalent area of the $i$th voxel, and $\rho_i$ represents the diffuse reflectance coefficient of the $i$th voxel.
Following the strategy of 'Simplicity before complexity,' this study decomposes the cylinder into two undersides and one lateral face to separately calculate their equivalent areas.
Based on empirical values, the lateral diffuse reflection coefficient is taken as 0.33, while the coefficients for two undersides are set to 0.03 and 0.02, respectively.

The formulas for calculating the equivalent area $A_{e,p}$ for planar voxels and $A_{e,c}$ for cylindrical voxel are as follows:
\begin{equation}
    A_{e,p} = A_p\cos{\alpha_p}\cos{\beta_p}
\end{equation}
\begin{equation}
    A_{e,c} = \frac{A_c\sin{\alpha_c}\sin{\beta_c}}{4\pi}[\sin{\gamma} + (\pi-\gamma)\cos{\gamma}]
\end{equation}
where $\gamma$ satisfies:
\begin{equation}
    \cos{\gamma} = \frac{\cos{\Phi}-\cos{\alpha_c}\cos{\beta_c}}{\sin{\alpha_c}\sin{\beta_c}}
\end{equation}
$A_p$ and $A_c$ represent the actual areas of the planar and lateral surfaces, with their respective normal directions pointing outward. The angles between the normal directions and the sun's position are denoted as $\alpha_p$ for the plane and $\alpha_c$ for the lateral voxel. Similarly, the angles between these directions and the observer's position are $\beta_p$ for the plane and $\beta_c$ for the lateral voxel. 

By substituting the formulas for equivalent areas into equation(\ref{equ: 5}), the photometry model of the target can be obtained shown as equation(\ref{equ: 9}), thereby calculating the theoretical light curves with equation(\ref{equ: 10}).
\begin{equation}\label{equ: 9}
    E_s = \frac{E_{0, V}}{\pi r^2}(\rho_1 A_{e,c} + \rho_2 A_{e,p\_bot} + \rho_3 A_{e,p\_top})
\end{equation}
\begin{equation}\label{equ: 10}
    m = -2.5 lg E_s + m_0
\end{equation}

\subsection{Lomb-Scargle periodogram}
The Lomb–Scargle periodogram (\citealt{Lomb+1976}, \citealt{Scargle+1982}) is a well-known algorithm for detecting and characterizing periodicity in unevenly sampled time series and has seen particularly wide use within the astronomy community (\citealt{VanderPlas+2018}). It serves as a powerful tool for analyzing irregularly sampled data and is applicable to various types of periodic signals. In the analysis of periodicity in light curves, the Lomb-Scargle algorithm is widely utilized to unveil the periodic behavior of targets.The Lomb-Scargle periodogram power calculation formula is shown below (\citealt{Huang+Ji+2016}):
\begin{equation}
    P_X(\omega)=\frac{1}{2}\left\{\frac{\left[\sum_j X_j \cos \omega\left(t_j-\tau\right)\right]^2}{\sum_j \cos ^2 \omega\left(t_j-\tau\right)}+\frac{\left[\sum_j X_j \sin \omega\left(t_j-\tau\right)\right]^2}{\sum_j \sin ^2 \omega\left(t_j-\tau\right)}\right\}
\end{equation}
$t_j$ denotes the time, $X_j$ represents the time series signal, and $P_X(\omega)$ stands for the Lomb-Scargle periodogram power. $\tau$ is calculated using the following formula:
\begin{equation}
    \tan (2 \omega \tau)=\frac{\sum_j \sin 2 \omega t_j}{\sum_j \cos 2 \omega t_j}
\end{equation}

In this study, the Lomb-Scargle algorithm is implemented by calling the \texttt{LombScargle} from the Python open-source library \texttt{Astropy} to extract the periods implied by the light curves. The \texttt{LombScargle} class is a unified interface to several implementations of the Lomb-Scargle periodogram.

\subsection{Automatic Differentiation algorithm and the loss function}

Automatic Differentiation is a powerful computational technique for evaluating the derivatives of functions specified by a computer program. It operates by systematically decomposing a function into a sequence of elementary arithmetic operations and standard functions (e.g., addition, multiplication, exponential functions). The chain rule is then applied recursively to these operations to compute the derivative value accurately and efficiently.

AD offers significant advantages over traditional approaches. Unlike symbolic differentiation, which can generate complex expressions, or numerical differentiation via finite differences, which is prone to truncation and round-off errors, AD calculates derivatives with machine precision. This combination of accuracy and efficiency has led to its widespread adoption, with many software packages providing seamless AD implementations (\citealt{Margossian+2019}). AD is generally implemented in two modes: forward and reverse.

Forward Mode: In the forward mode, differentiation starts from the input variables and proceeds sequentially through each node of the computation graph in the forward direction until the derivative of the final objective function is obtained. 

Reverse Mode: In the reverse mode, differentiation starts from the output variables and proceeds sequentially through each node of the computation graph in the reverse direction until the derivative of the initial input variables is obtained. In this study, the open-source package \texttt{Jax} (\citealt{Bradbury+etal+2021}) is employed to implement the AD algorithm.

To illustrate the mechanism, Table~\ref{tab:1} provides a step-by-step example of the forward mode AD process for a sample expression~\ref{eq:14}. Given initial inputs of $x_a = 2$ and $x_b = 5$, the table traces the evaluation of each intermediate variable and its corresponding derivative with respect to $x_a$.

\begin{equation} \label{eq:14}
    f(x_a, x_b) = \ln(x_a) + x_ax_b - \sin(x_b)
\end{equation}
\begin{table}
    \renewcommand{\arraystretch}{1.5} 
	\begin{tikzpicture}[overlay,remember picture]
	\draw[->,thick] (-3.3,-0.7) -- (-3.3,-6.8); 
    \draw[->,thick] (-3.3,-7.7) -- (-3.3,-13.8); 
	\end{tikzpicture}
    \centering
    \setlength{\arrayrulewidth}{1.1pt} 

    \begin{tabular}{ll}
	\hline
	\textbf{Forward evaluation trace}   \\
	\ \ $v_a = x_a $      & $= 2$  \\
	\ \ $v_b = x_b $      & $= 5 $ \\
    \hdashline[1pt/2pt]
	\ \ $v_1 = lnv_{a}  $& $=ln2 $ \\
    \ \ $v_2 = v_{a} \times v_b  $& $= 2 \times 5 $ \\
    \ \ $v_3 = \sin{v_b}$  & $= \sin{5}$ \\
    \ \ $v_4 = v_1 + v_2$ & $= 0.693 + 10$ \\
    \ \ $v_5 = v_4 - v_3$ & $= 10.693 + 0.959$ \\
    \hdashline[1pt/2pt]
    \ \ $y = v_5$ & $= 11.652$\\
    \hline

    \textbf{Forward derivative trace } \\
    \ \ $\dot{v}_a = \dot{x}_a$ &  $= 1$\\
    \ \ $\dot{v}_b = \dot{x}_b$ &  $= 0$\\
    \hdashline[1pt/2pt]  
    \ \ $\dot{v}_1 = \dot{v}_a/v_a$ &  $= 1/2$\\
    \ \ $\dot{v}_2 = \dot{v}_a \times v_b +\dot{v}_b \times v_a$ & $= 1 \times 5 + 0 \times 2$\\
    \ \ $\dot{v}_3 = \dot{v}_b \times \cos{v_b}$ &  $= 0 \times \cos{5}$\\
    \ \ $\dot{v}_4 = \dot{v}_1 + \dot{v}_2$ &  $= 0.5 + 5$\\
    \ \ $\dot{v}_5 = \dot{v}_4 - \dot{v}_3$ & $= 5.5 - 0$\\
    \hdashline[1pt/2pt]
    \ \ $\boldsymbol{\dot{y} = \dot{v}_5}$ & $\boldsymbol{= 5.5}$\\
    \hline
    \end{tabular}
    
	\caption{Forward-mode AD evaluation and derivative traces for the function $f(x_a, x_b)$. $\dot{v}_i$ satisfies $\dot{v}_i = \frac{\partial v_i}{\partial x_a}$ and $\dot{y}$ satisfies  $\dot{y} = \frac{\partial y}{\partial x_a}$.}
	\label{tab:1}
	\end{table}

To optimize the attitude parameters, we define a loss function that quantifies the discrepancy between the observed light curve and the theoretical light curve generated by our photometric model. We employ the MSE, a standard and effective metric for this fitting task.
For a light curve consisting of $n$ observations, let $O_j$ be the measured magnitude at time $t_j$, and let $M_j(t_j, \lambda, \theta, \phi_0, T)$ be the corresponding theoretical magnitude computed by our model for a given set of attitude parameters $(\lambda, \theta, \phi_0)$. The MSE loss function, $L$, is then defined as the average of the squared residuals:

\begin{equation}\label{equ: 14}
    L = \frac{1}{n} \sum_{j = 1}^n [M_j(t_j, T, \lambda, \theta, \phi_0) - O_j]^2
\end{equation}
The objective of our method is to find the set of attitude parameters $(\lambda, \theta, \phi_0)$ that minimizes the loss function, $L$. To accomplish this, we employ a gradient descent optimization algorithm. The critical step in this process is the computation of the gradient of the loss function with respect to the attitude parameters. We compute this gradient efficiently and with machine precision using AD. The parameters are then iteratively updated until the loss converges to a minimum. This approach replaces the computationally expensive grid search of prior work with an efficient optimization, enabling rapid and precise attitude estimation.


\section{Results}
\label{sect:result} 

\subsection{Lomb-Scargle algorithm}
\begin{figure}
\centering
	\includegraphics[width=12cm, keepaspectratio]{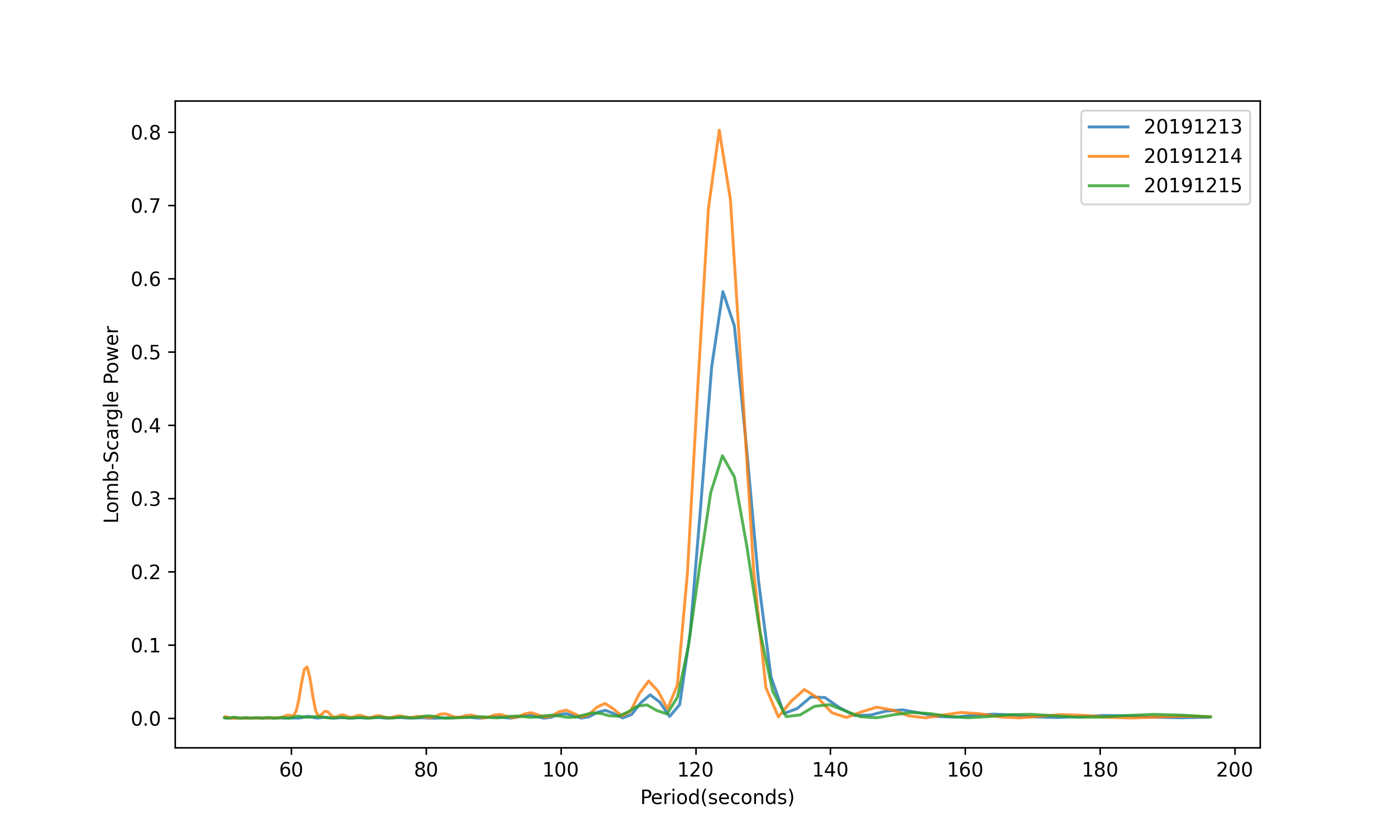}
    \caption{Lomb–Scargle periodogram computed from the light curve data.}
    \label{fig:7}
\end{figure}

The rotation period of the target was determined using the Lomb-Scargle periodogram method. Figure~\ref{fig:7} shows the resulting power spectrum for the light curves from each of the three nights of observation. The horizontal axis represents the period, and the vertical axis indicates the power. The analysis reveals prominent peaks in the power spectrum at 124.045 s, 123.510 s, and 123.975 s for the three respective nights. Due to the cylindrical symmetry of the target, the photometric period is expected to be half of the true rotational period. Therefore, the actual rotation periods are determined to be 248.091 s, 247.020 s, and 247.950 s.

\begin{figure}
	\includegraphics[width=\columnwidth]{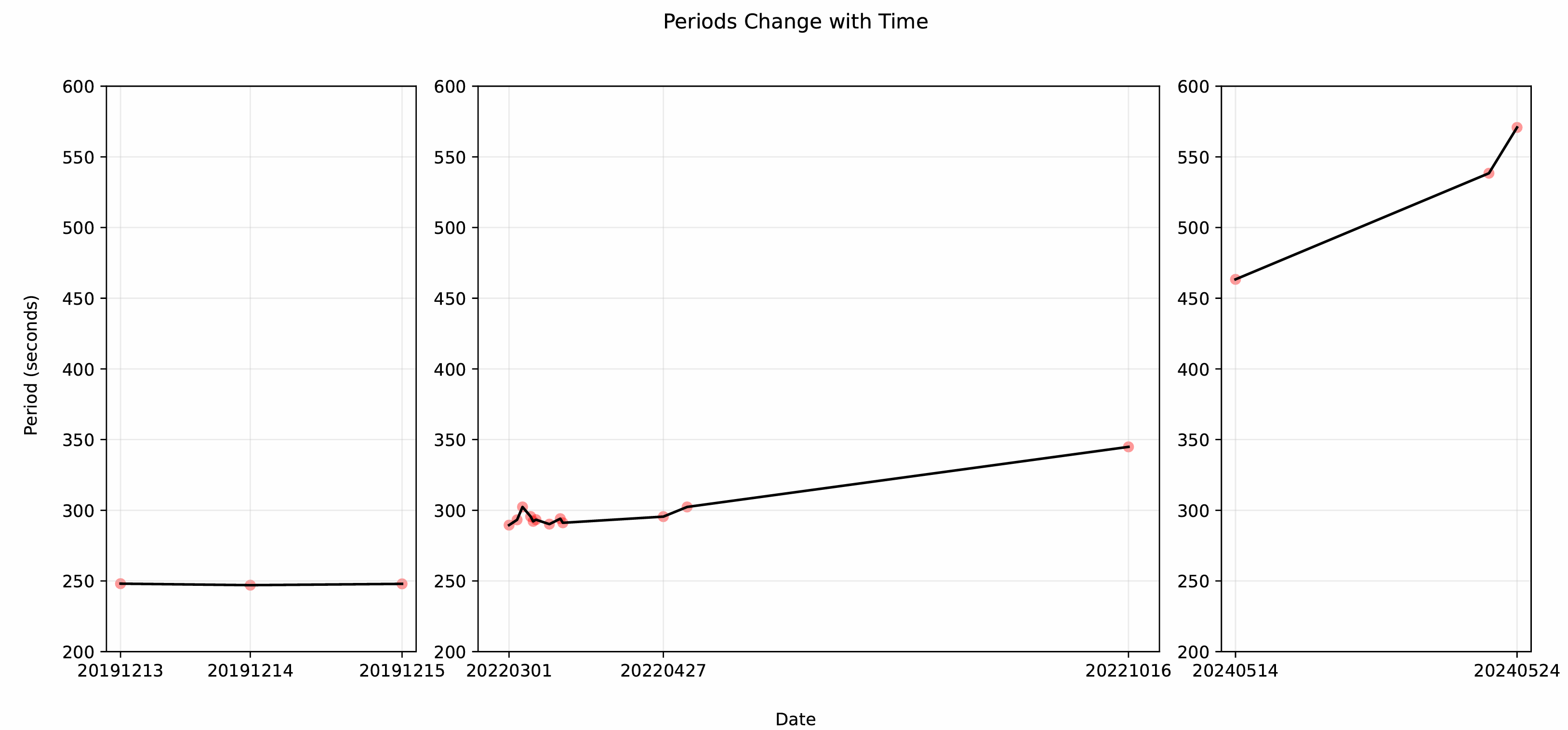}
    \caption{The rotational periods of the target vary over time, with the horizontal axis rep
resenting the date and the vertical axis denoting the rotational periods (in seconds).}
    \label{fig:8}
\end{figure}

This analysis was extended to light curves from additional observation dates, with the resulting periods plotted over time in Figure~\ref{fig:8}. The figure reveals a clear secular increase in the rotation period, indicating a gradual spin-down of the target. This trend is consistent with the effects of dissipative environmental torques.

\subsection{Attitude parameters optimization using the AEAD algorithm}

\subsubsection{Validation of algorithm effectiveness based on simulated light curves} 
We first validated the AEAD algorithm using the two simulated light curves shown in Figure~\ref{fig:3}. These simulations represent observations taken several days apart, providing diverse viewing geometries, which is crucial for determining the attitude.

\begin{figure}
    \centering
        \includegraphics[width=10cm, keepaspectratio]{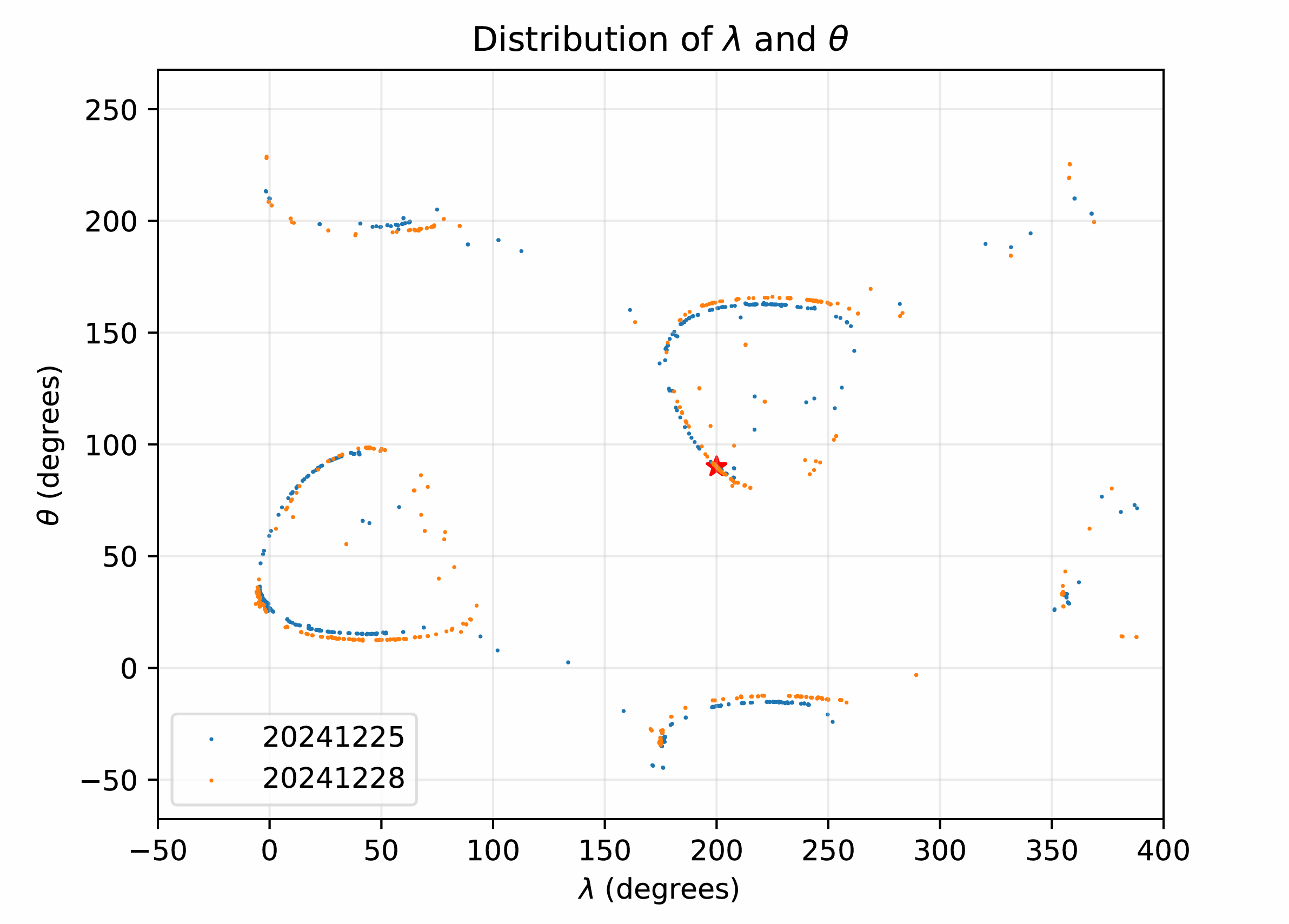}
        \caption{The distributions of the optimal rotation axis orientations in the $\lambda - \theta$ plane, obtained by applying AEAD method and performing a 
        grid search on the simulated light curves for the dates 2024-12-25 and 2024-12-28, are presented. The red pentagram indicates the true rotation axis orientation 
        for the simulated light curves.}
        \label{fig:16}
\end{figure}

Applying AEAD to each light curve independently yields a family of possible solutions, which trace circular paths in the parameters space, as shown in Figure~\ref{fig:16}. This ambiguity arises because a single light curve cannot distinguish between different spin axis orientations that produce the same viewing aspect over a rotation. The red pentagram in the figure marks the ground-truth orientation of (200, 90).

\begin{figure}
    \centering
        \includegraphics[width=10cm, keepaspectratio]{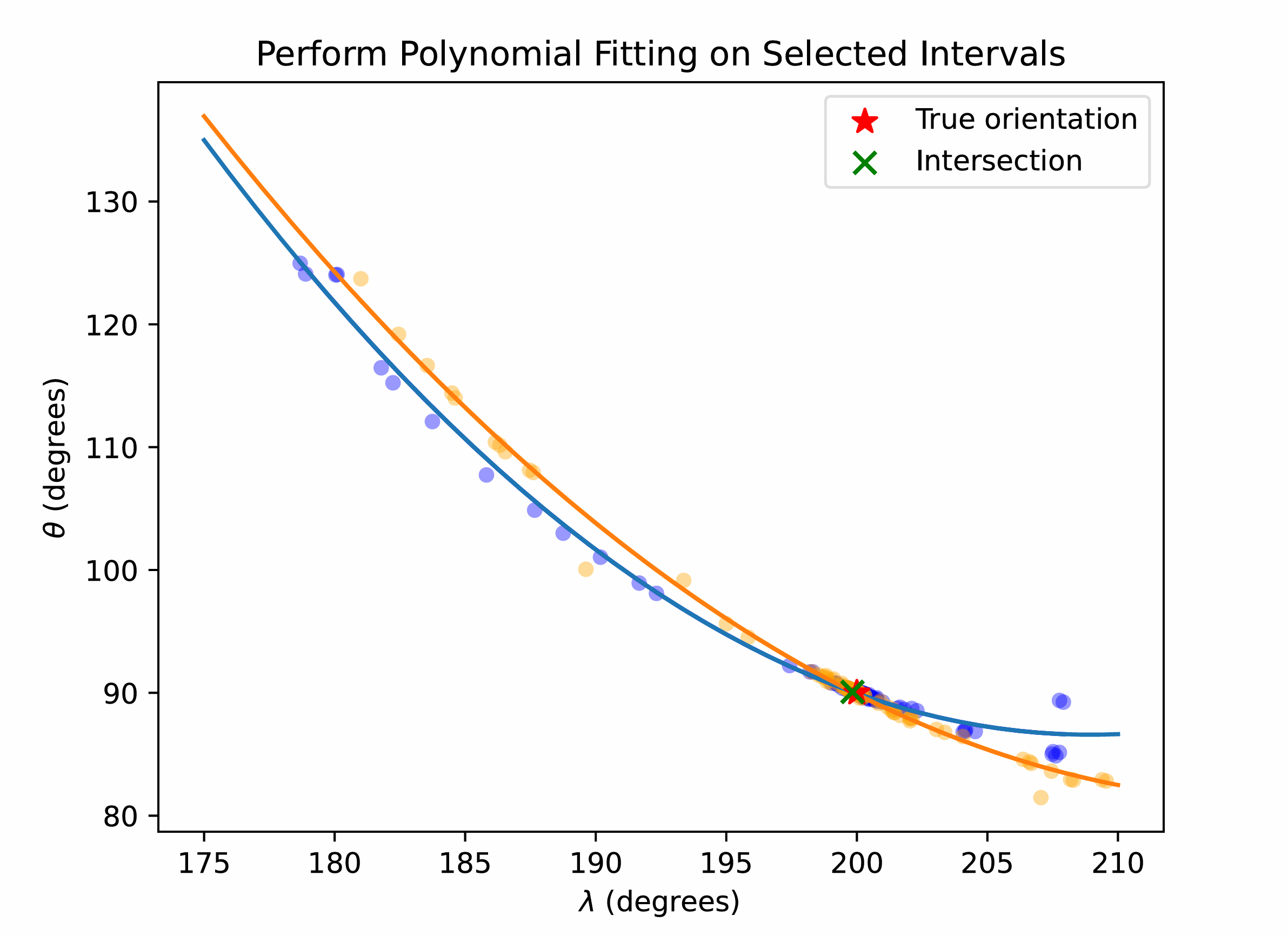}
        \caption{Perform Polynomial fittings on selected intervals of the circular distribution. 
        The yellow and blue points are extracted from specific intervals in Fig.~\ref{fig:16}, while the yellow and blue curves represent the polynomial 
        fits to these points. The green cross marker, located at (199.832, 90.073), indicates the intersection of the fitted curves, and the red pentagram, 
        positioned at (200, 90), denotes the true rotation axis orientation of the simulated light curves.}
        \label{fig:17}
\end{figure}

This ambiguity is resolved to a large extent by combining the constraints from both observation geometries. The true spin axis must lie at the intersection of the two solution curves. To find this point, we perform a polynomial fit to each distribution (Figure~\ref{fig:17}) and calculate their intersection. The calculated intersection at (199.832, 90.073) represents our final estimate of the spin axis orientation. This result demonstrates good agreement with the ground-truth value, confirming the high accuracy of our method.


\subsubsection{Attitude estimation based on observed light curves from YNAO} 
The previous study by \cite{Su+etal+2021} relied on an exhaustive grid search to find the optimal attitude parameters. Their approach required discretizing the parameters space—[0, 360] for $\lambda$, [0, 180] for $\theta$, and [0, 360] for $\phi_0$—with step sizes of 2, 1, and 10 degrees, respectively. Consequently, there are over one million possible combinations for these three parameters. Although this method identified two sets of parameters that minimize the MSE, it was highly time-consuming and inefficient. 

\begin{figure}
\centering
	\includegraphics[width=10cm, keepaspectratio]{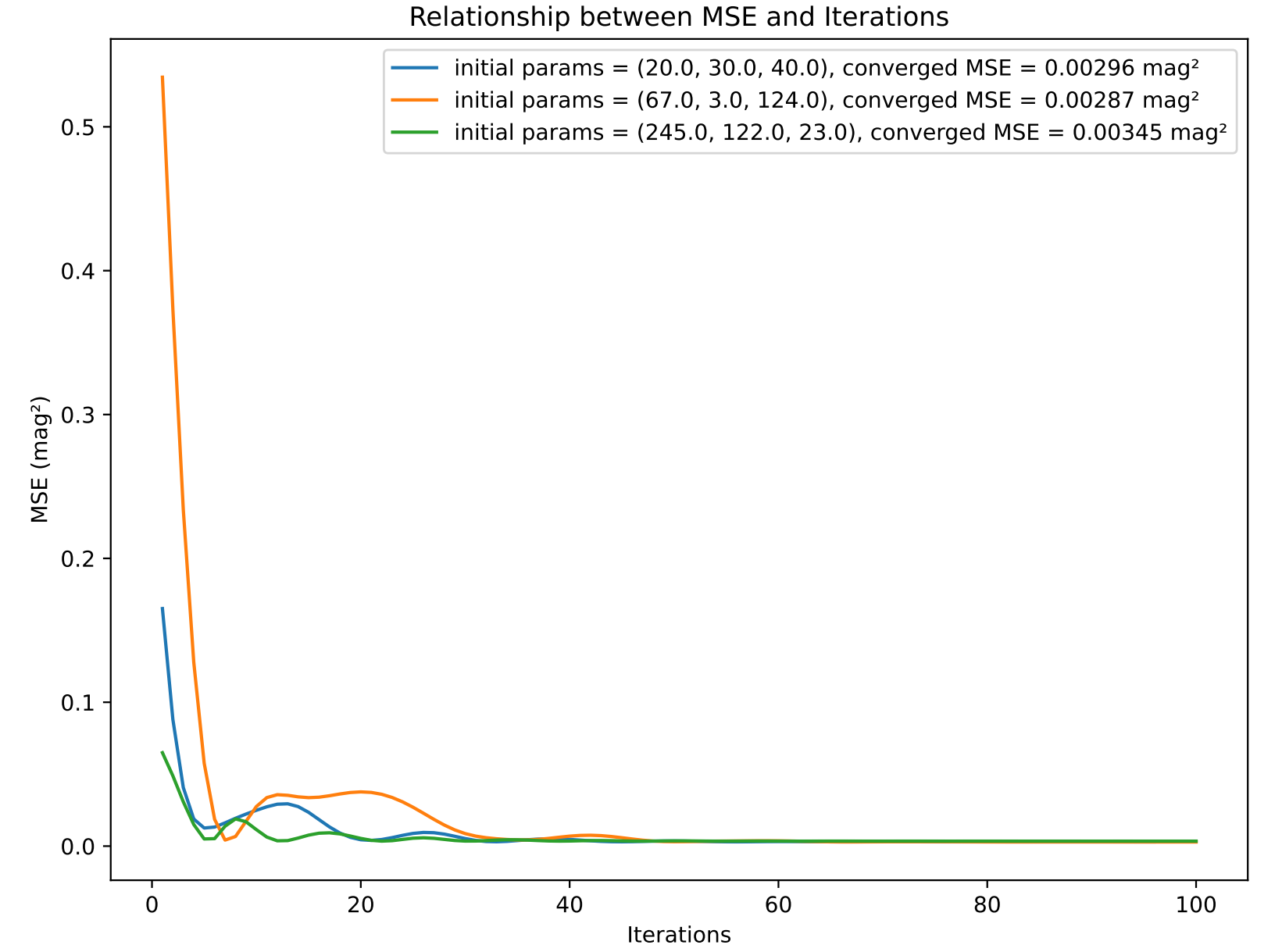}
    \caption{Using the light curve data from 20191213, as an example, the AEAD method can estimate parameters efficiently. Regardless of the initial parameters, convergence to a stable solution is achieved within approximately 50 iterations (about 10 seconds). Moreover, the MSEs can be reduced to three decimal places. This algorithm exhibits similar performance with other dates.}
    \label{fig:AEAD方法迭代情况}
\end{figure}

\begin{figure}
	\includegraphics[width=\columnwidth]{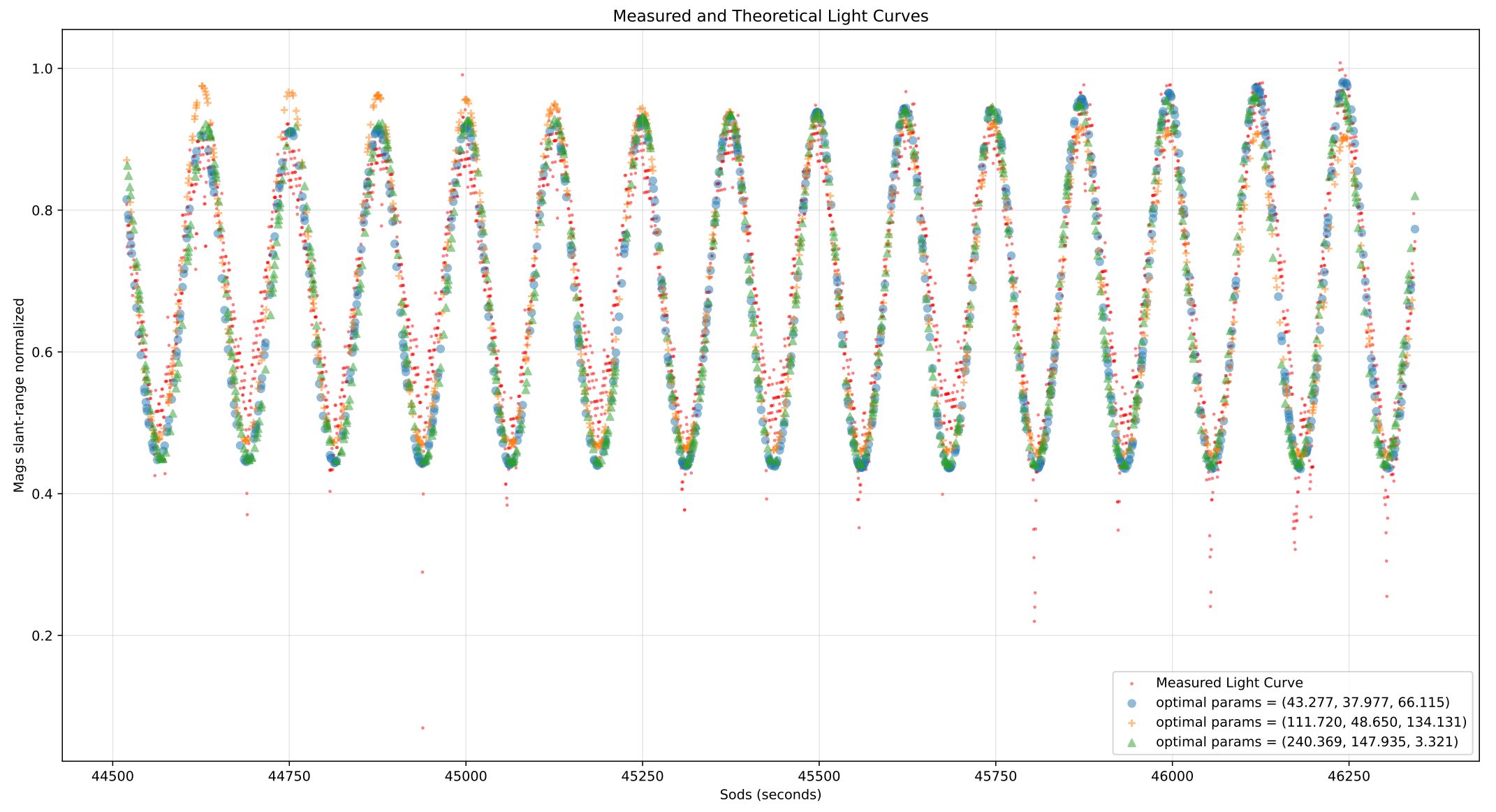}
    \caption{From the three sets of random initial parameters shown in Fig.~\ref{fig:AEAD方法迭代情况}, three sets of optimal parameters are obtained through convergence. Applying these parameters to calculate the theoretical light curves, we find good agreements with the observed data. There are some outliers at the peaks of the magnitude, which are caused by the target's specular reflection.}
    \label{fig:三组参数的理论光变曲线符合很好}
\end{figure}

In stark contrast, our AEAD method provides a vastly more efficient solution. As demonstrated in Figure~\ref{fig:AEAD方法迭代情况}, our algorithm consistently converges to an optimal solution in approximately 50 iterations (taking about 15 seconds), even when starting from random initial parameters. The resulting theoretical light curves provide an excellent fit to the observational data, as shown in Figure~\ref{fig:三组参数的理论光变曲线符合很好}. Crucially, our approach achieves a final Mean Squared Error (MSE) of approximately 0.003, which represents an order-of-magnitude improvement in fit quality compared to the previous work.


During our analysis of the 20191213 light curve, we observed that the AEAD algorithm converged to different optimal solutions depending on the initial parameters. We hypothesized that this multiplicity of solutions is a direct consequence of the target's geometric symmetry. To investigate this, we systematically explored the solution space by running the optimization from a grid of initial parameters values.

\begin{figure}
\centering
	\includegraphics[width=10cm, keepaspectratio]{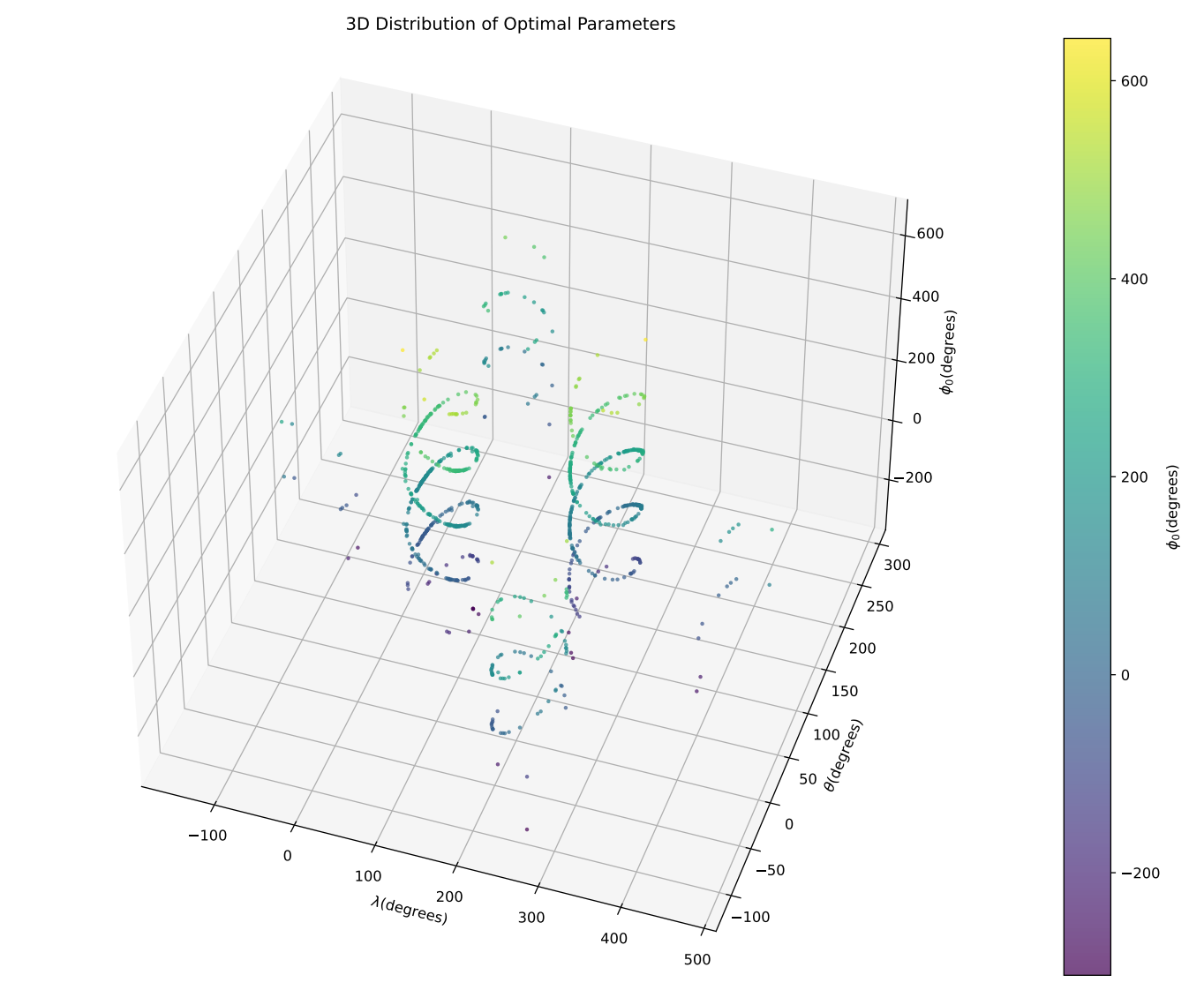}
    \caption{After discretizing the initial parameters into a grid, parameters estimation is performed on all combinations, resulting in optimal parameters sets $\lambda, \theta, \phi_0 $, exhibiting a double helical distribution in the inertial coordinate system. $\phi_0$ were mapped onto a color scale representing heat map.}
    \label{fig: 参数的三维分布}
\end{figure}

\begin{figure}
\centering
	\includegraphics[width=10cm, keepaspectratio]{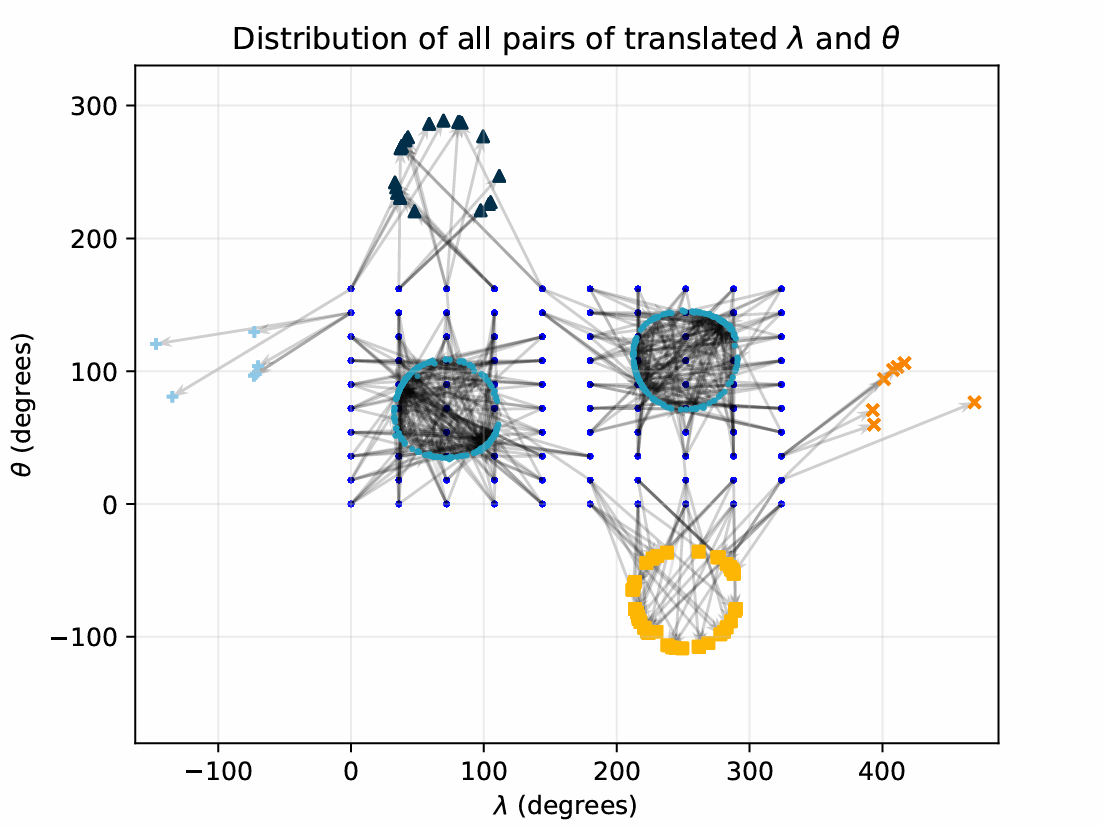}
    \caption{By projecting Fig.~\ref{fig: 参数的三维分布} onto a plane, it can be observed that $\lambda$ and $\theta$ exhibit circular distributions which are represented using points with varying colors and shapes in different regions of the plane. Blue points denote the initial parameters combinations, while points of other colors represent the corresponding optimization results. Gray arrows are used to connect the initial parameters and the optimal rotation axis orientations, illustrating their correspondence.}
    \label{fig: 参数的二维分布}
\end{figure}

Leveraging the high efficiency of the AEAD method, we performed a coarse grid search for the initial values of $\lambda$, $\theta$, and $\phi_0$ with step sizes of 36, 18, and 36 degrees, respectively. This comprehensive search revealed a distinct and structured set of optimal parameters combinations. When plotted in 3D space, these solutions form a striking double helix pattern, as shown in Figure~\ref{fig: 参数的三维分布}. The projection of these solutions onto the 2D ($\lambda$, $\theta$) plane forms a roughly circular distribution, as illustrated in Figure~\ref{fig: 参数的二维分布}.


\begin{figure}
\centering
	\includegraphics[width=10cm, keepaspectratio]{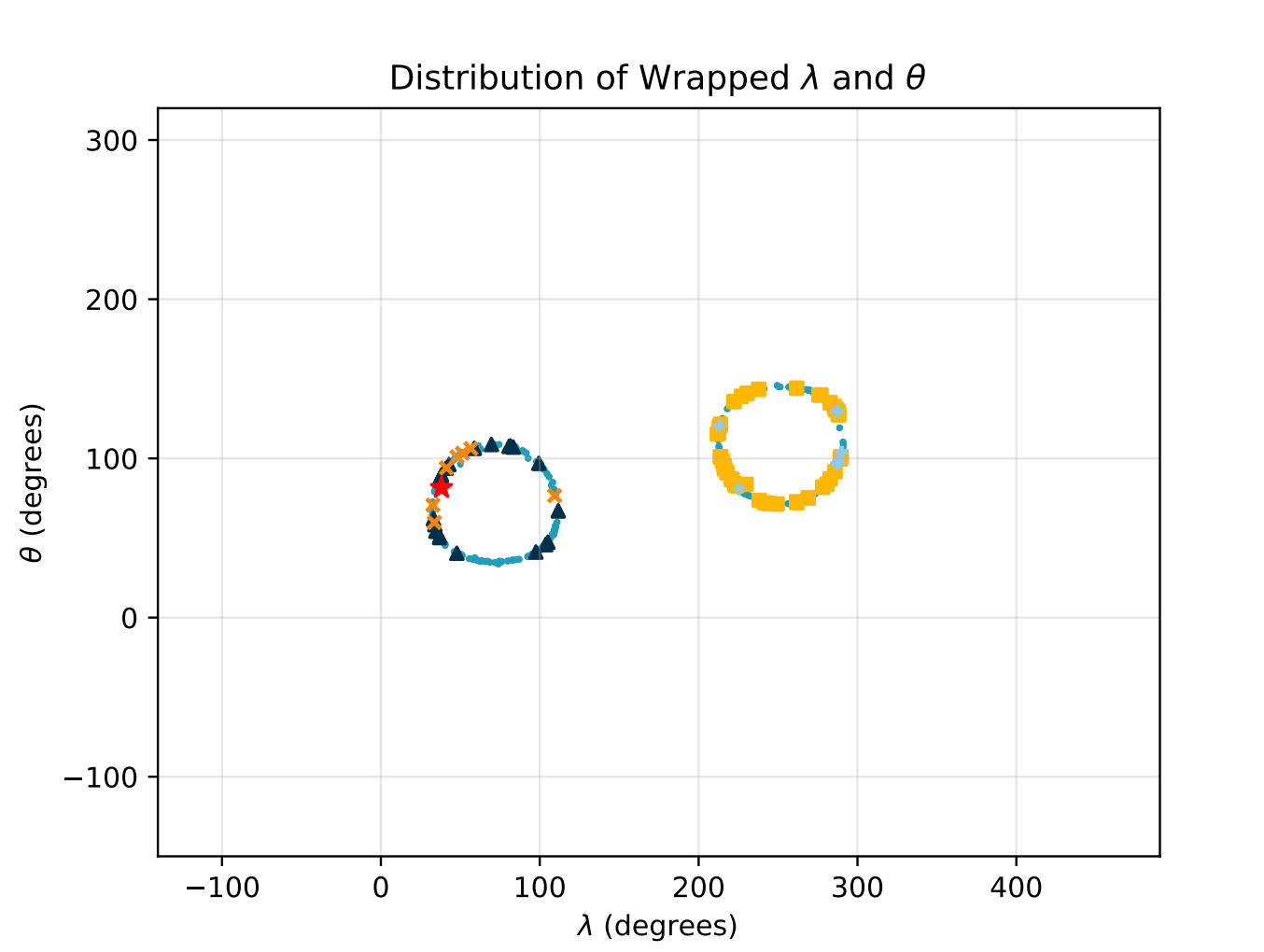}
    \caption{By wrapping the circles in Fig.~\ref{fig: 参数的二维分布} along the coordinate axis by either 180 degrees or 360 degrees, they can be superimposed to form two overlapping circles. The reason behind the formation of these two circles is the positive and negative directions of the rotation axis.}
    \label{fig:13}
\end{figure}

The observed distributions of the optimal parameters in Figure~\ref{fig: 参数的二维分布} are a direct result of their periodic nature. Given the 180° and 360° periodicities in the attitude parameters, the multiple solution sets can be mapped onto two fundamental solution circles by applying appropriate coordinate shifts, as demonstrated in Figure~\ref{fig:13}.

\begin{figure}
\centering
	\includegraphics[width=10cm]{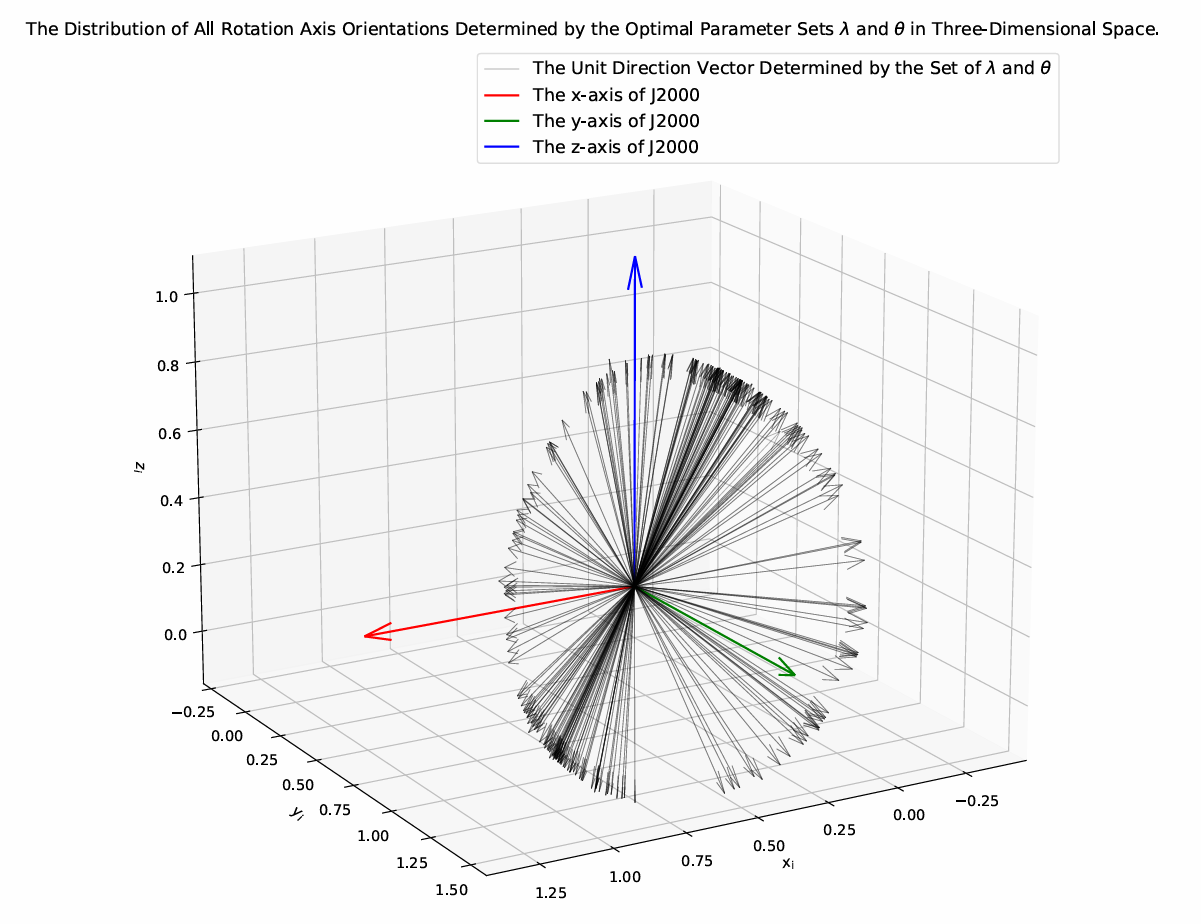}
    \caption{In the J2000 coordinate system, the directions of the rotation axis determined by the optimal parameters $\lambda$ and $\theta$ are indicated by a black arrows. The rotational symmetry of the rotation axis direction indicates the presence of the axis of rotational symmetry.}
    \label{fig:14}
\end{figure}

The existence of these two distinct circles can be explained by a fundamental symmetry in the model: for any spin-axis solution, an equally solution is obtained by reversing the direction of the rotation axis (i.e., pointing it in the opposite direction). This corresponds to one solution circle being the antipode of the other. This inherent symmetry in the solution space is further visualized in three dimensions in Figure~\ref{fig:14}, which illustrates the rotational symmetry of the spin-axis solutions.

\begin{figure}
\centering
	\includegraphics[width=8cm, keepaspectratio]{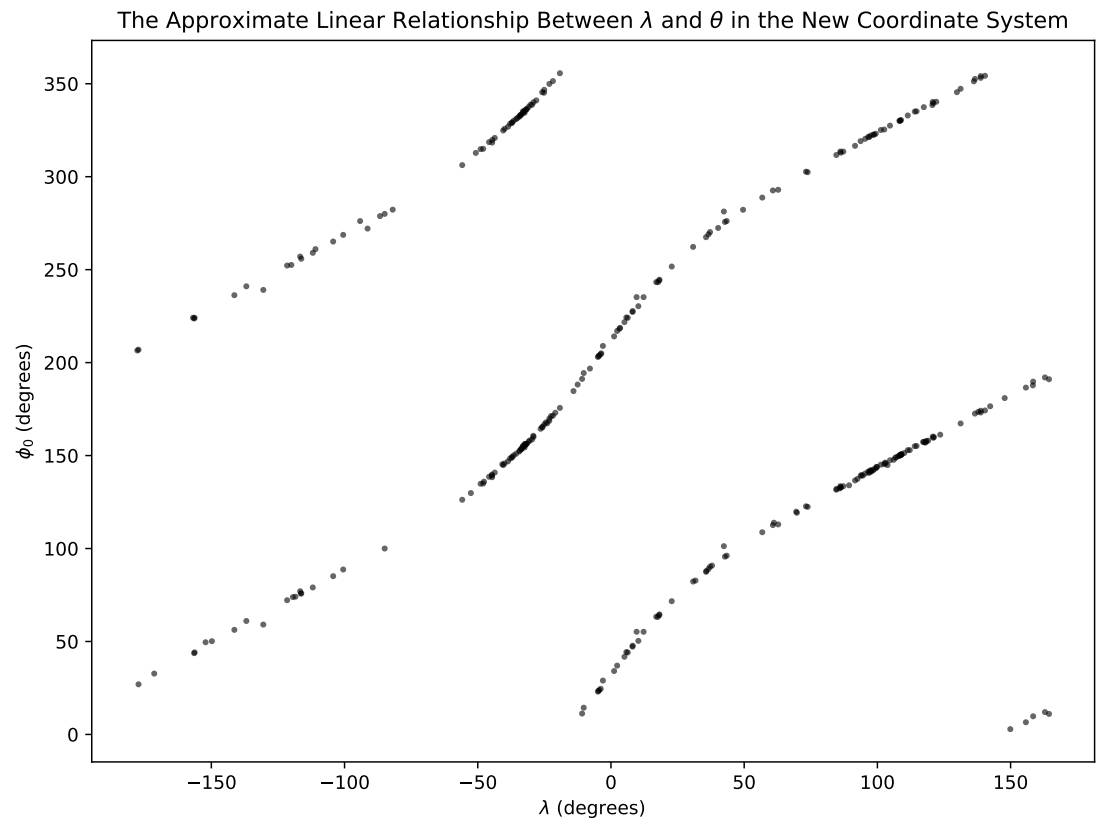}
    \caption{In the new coordinate system, $\lambda$ and $\phi_0$ exhibit approximately linear distribution.}
    \label{fig:lambda_vs_phi0}
\end{figure}

After rotating the inertial coordinate system to align the z-axis with the axis of rotational symmetry, the spin-axis solutions exhibit symmetric distribution around the z-axis in the new coordinate system. In this new coordinate system, it can be observed that $\lambda$ and $\phi_0$ show approximately linear distribution, as depicted in Fig.~\ref{fig:lambda_vs_phi0}.

The efficiency of the AEAD method not only accelerates the primary task of attitude estimation but also enables the exploration of additional model parameters. We extended our approach to simultaneously estimate the effective diffuse reflectance coefficient, $\rho_i$, for each of the cylinder's surface components, in addition to the attitude parameters $\lambda$, $\theta$, and $\phi_0$.

As shown in Table~\ref{tab:2}, the AEAD algorithm successfully converges to a set of $\rho_i$ values that yield a low MSE, demonstrating its capability for model parameters inversion. This result is significant, as it provides a data-driven method for determining surface properties, which can replace or refine the empirically chosen values often used in photometric models.

\begin{table}
\centering
\begin{tabular}{cccc}
\toprule
Date & Initial $\rho_1,\ \rho_2,\ \rho_3$ & Optimal $\rho_1,\ \rho_2,\ \rho_3$ & MSE \\
\midrule
2019-12-13 & 0.340, 0.010, 0.010 & 0.326, 0.054, 0.048 & 0.002 \\
2022-03-10 & 0.340, 0.010, 0.010 & 0.333, 0.244, 0.149 & 0.003 \\
2024-05-23 & 0.340, 0.010, 0.010 & 0.339, 0.012, 0.143 & 0.006 \\
\bottomrule
\end{tabular}
\caption{Apply AD for parameters estimation of the effective diffuse reflectance coefficient. $\rho_1$, $\rho_2$, and $\rho_3$ represent the coefficient of the lateral face and two undersides, respectively. The second column in the table shows the initial params, while the third column presents the optimized results.}
\label{tab:2}
\end{table}

\subsubsection{Attitude estimation based on observed light curves from MMT} 
To test the generalizability of our algorithm, we applied the AEAD method to an independent dataset: the observed light curves of the same target obtained by the MMT system on October 26 and October 29, 2019 (presented in Figures~\ref{fig:4} and~\ref{fig:5}).

\begin{figure}
    \centering
        \includegraphics[width=10cm, keepaspectratio]{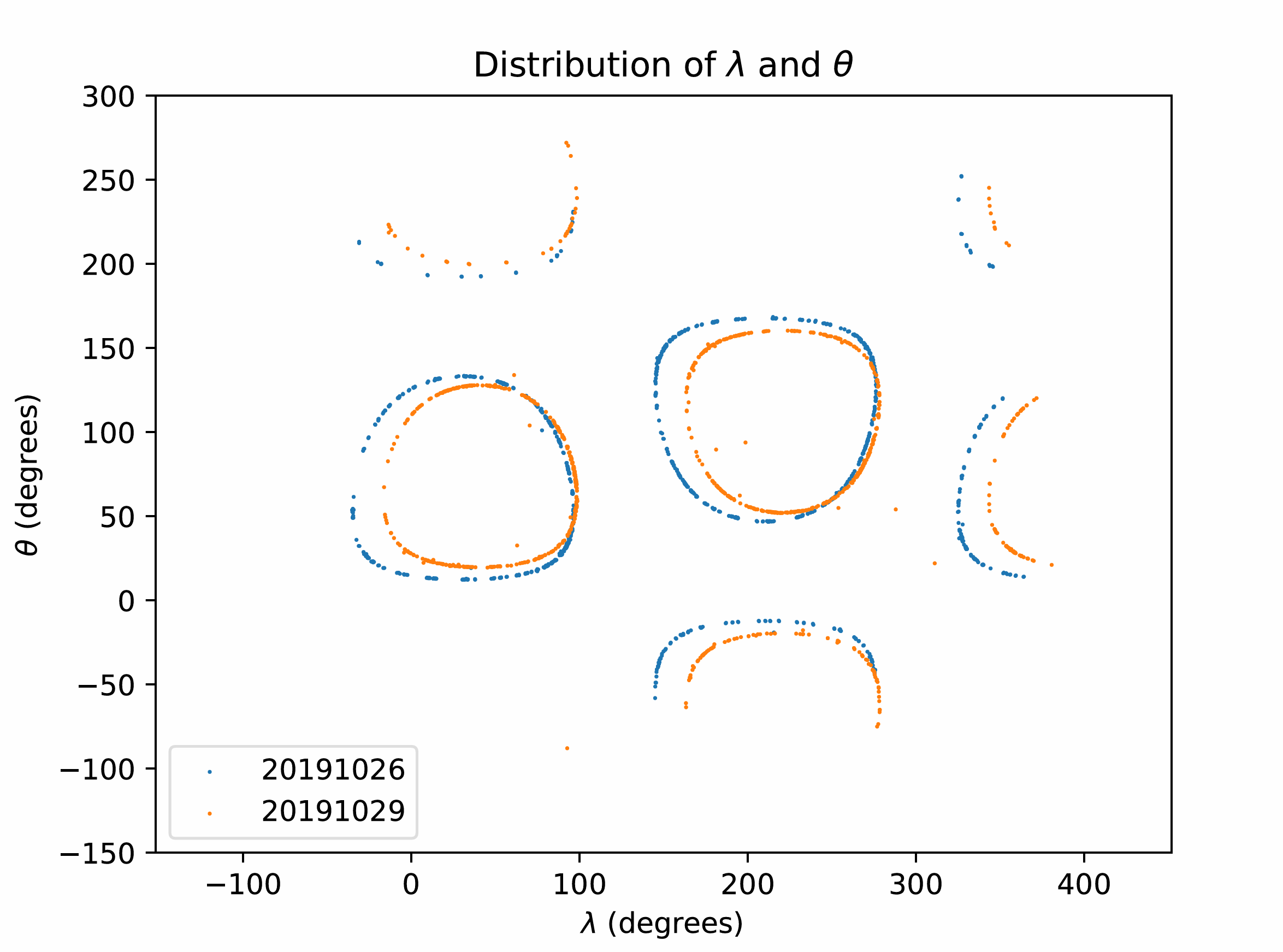}
        \caption{Apply the AEAD method to two light curves from MMT and conduct a grid search.}
        \label{fig:18}
\end{figure}

The analysis of the MMT data yielded results that are highly consistent with our primary findings. Specifically, the optimal solutions for the spin axis orientation form the same characteristic circular distributions, as shown in Figure~\ref{fig:18}. Successfully replicating these results on a different dataset, from a different instrument, provides strong evidence for the robustness and generalization capability of the proposed AEAD method.

\subsubsection{Analysis of attitude variation based on dual-site observed light curves}
\begin{figure}[!htbp]
    \centering
    \includegraphics[width=0.8\textwidth]{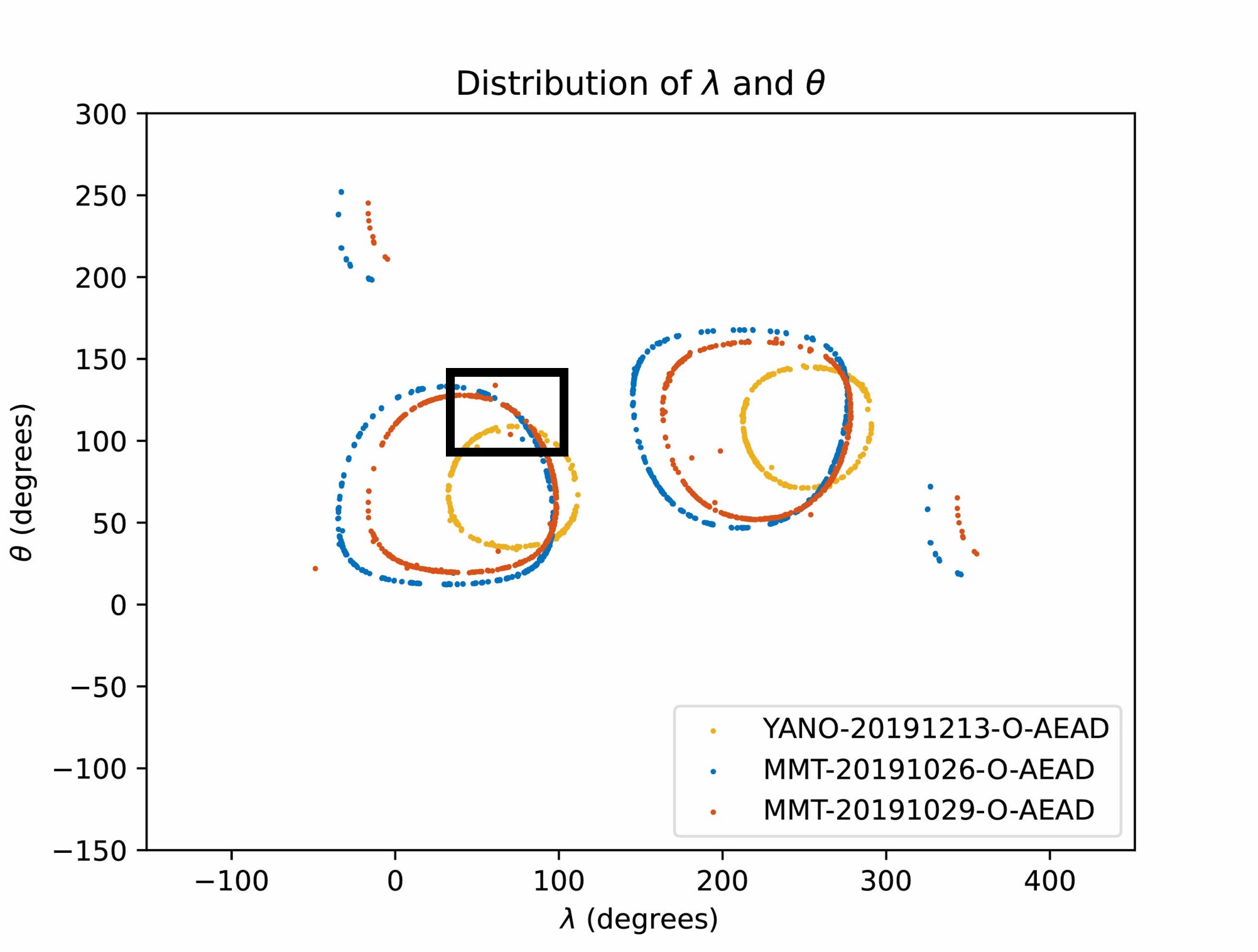}
    \caption{The optimal parameters distribution obtained by applying the AEAD algorithm to three days of observed light curves from YNAO and MMT.}
    \label{三天实测-AEAD}
\end{figure}

A combined analysis of the attitude constraints from both the YNAO and MMT datasets provides robust determination of the spin axis orientation. As illustrated in Figure~\ref{三天实测-AEAD}, the solution circles derived from the YNAO data (yellow) and the MMT data (red and blue) intersect in a common region of the parameters space.

\begin{figure}[!htbp]
    \centering
    \includegraphics[width=0.8\textwidth]{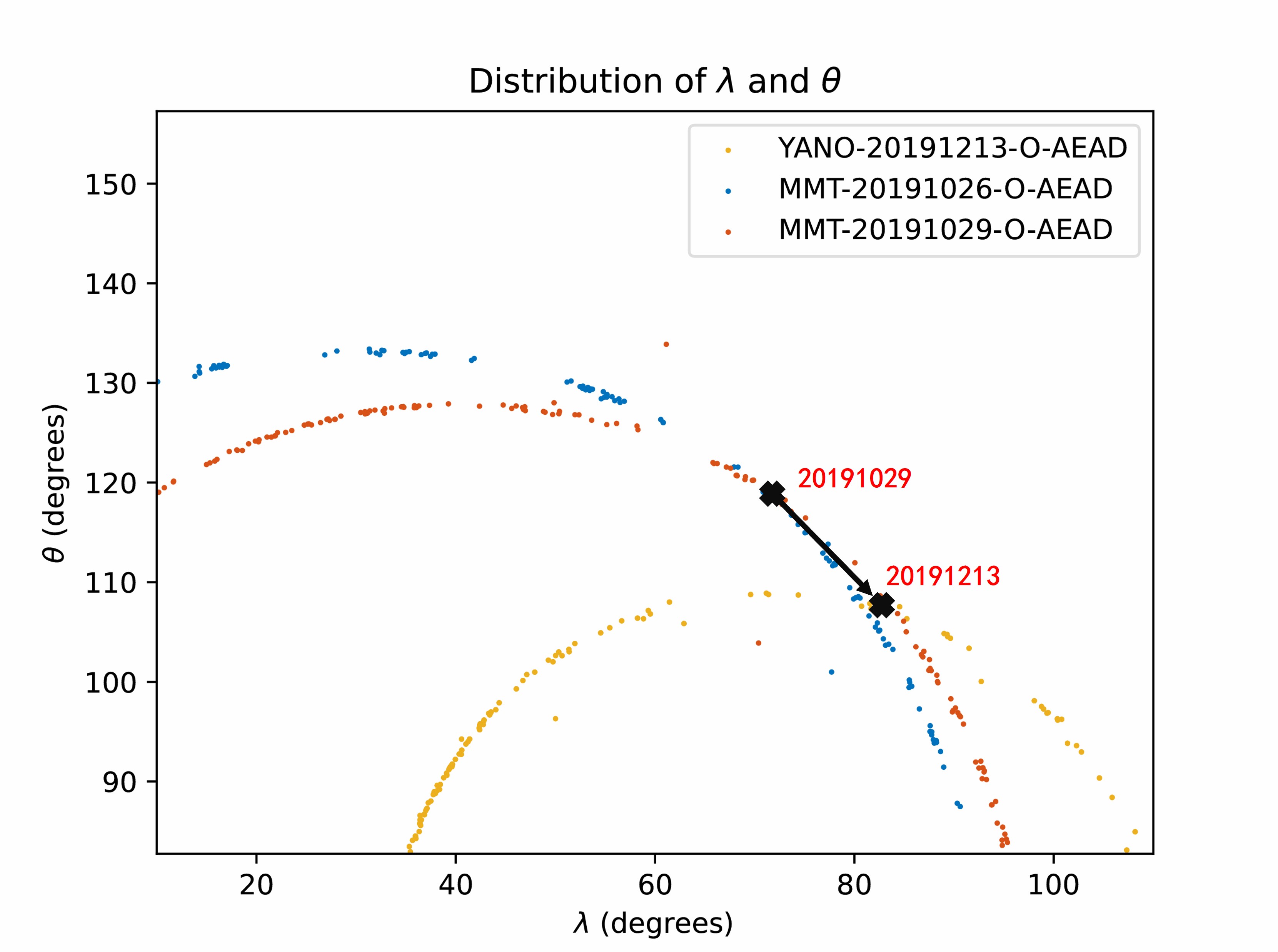}
    \caption{Magnified view of the region outlined by the black rectangle in Fig~\ref{三天实测-AEAD}. The displayed area covers the parameter space of $\lambda \in [10, 110]^\circ$ and $\theta \in [85, 150]^\circ$.
    \label{指向演化}}

\end{figure}

Crucially, the intersection of these solution sets does not converge to a single, static point but instead traces a clear path, as highlighted in the magnified view (Figure~\ref{指向演化}). This systematic temporal evolution indicates that the target's spin axis is not fixed but is precessing over the observation period. This finding demonstrates the sensitivity of our method, allowing us to detect subtle dynamic effects that were beyond the scope of the initial, simplified model.

Figure~\ref{指向演化} provides a detailed view of this temporal evolution, allowing us to track the changes of the spin axis. The intersection of the two MMT solution circles (red and blue from late October) pinpoints the spin axis orientation at that epoch. Subsequently, the intersection of the YNAO data (yellow circle from December 13) with the earlier MMT constraints defines a new, distinct solution region.

The clear migration of this intersection point over about 45-day period between the observation sets provides a direct measurement of the target's spin axis change. This demonstrates that our method is not only capable of determining a static attitude but is also sensitive to characterize its dynamic evolution over time.
This capability to rapidly and accurately process data from different instruments at different times is what allowed us to perform a dynamic analysis of the target's rotational state. By successfully characterizing this temporal evolution, our work establishes a powerful new methodology for future research into the long-term rotational dynamics of space objects and the environmental torques that drive them.

\subsubsection{Performance comparison between grid search and AEAD algorithms} 
The AEAD algorithm developed in this study offers significant improvements in both efficiency and accuracy over the grid search method previously employed by \cite{Su+etal+2021}. This superior performance is quantitatively summarized in Table~\ref{tab:算法性能比较}, which highlights the key metrics of both approaches. The results demonstrate that AEAD drastically reduces the computation time required to find an optimal solution while simultaneously achieving a lower MSE, indicating a more accurate model fit. The expression for MSE is given in Equation~\eqref{equ: 14}.

Furthermore, our work confirms the robust generalization capability of the AEAD algorithm. Its effectiveness was validated across a diverse and challenging set of observations, including multi-year data spanning from 2019 to 2024 and multi-site data from two independent observatories (YNAO and MMT). This ability to rapidly and accurately process multi-source data is a key advantage, enabling the fusion of datasets required to detect subtle, long-term dynamic effects.

Nevertheless, the corresponding MSE values on the MMT data, in Table~\ref{tab:算法性能比较}, do not show significant improvement. This is because an error-absorption term is introduced in the photometric model primarily to compensate for systematic discrepancies under the specific observing conditions of YNAO, including differences between the actual target size, surface material properties, and the idealized model assumptions. These correction terms are calibrated using YNAO data and exhibit a certain degree of scene dependence. Therefore, the dominant error sources in the corresponding observation scenario can be effectively absorbed, leading to an improvement in modeling accuracy. In contrast, the MMT dataset originates from a different telescope system and observation geometry. Consequently, its dominant error sources, noise characteristics, and systematic biases may differ from those of the YNAO data. As a result, when applying the calibrated model to the MMT dataset, the same level of MSE improvement is not observed. 

\begin{table}[h]
    \centering
    \caption{Algorithms Performance Comparison}
    \small  
    \begin{tabular}{
        >{\raggedright\arraybackslash}p{1.8cm}  
        p{1.2cm}  
        p{1.5cm}  
        p{1.8cm}  
        p{1.6cm}  
        p{1.7cm}  
        p{0.8cm}  
    }
        \hline
        \textbf{Algorithm} & \textbf{Target} & \textbf{Station} & \textbf{Date} & \textbf{Data Type} & \textbf{Time Cost} & \textbf{MSE} \\
        \hline
        Grid Search
        & 42968 & YNAO
        & 2019-12-13 & Observed & ~24 hours & 0.046 \\
     
        \cline{1-7}
        
        \multirow{7}{*}{AEAD} 
        & \multirow{7}{*}{42968} & \multirow{5}{*}{YNAO} 
        & 2019-12-13 & Observed & \multirow{5}{*}{~15 seconds} & 0.003 \\
        & & & 2019-12-14 & Observed &  & 0.009 \\
        & & & 2019-12-15 & Observed &  & 0.002 \\
        & & & 2024-12-25 & Simulated &  & 0.004 \\
        & & & 2024-12-28 & Simulated &  & 0.005 \\
        \cline{3-7}
        & & \multirow{2}{*}{MMT} 
        & 2019-10-26 & Observed & \multirow{2}{*}{~15 seconds} & 0.041 \\
        & & & 2019-10-29 & Observed &  & 0.040 \\
        \hline
    \end{tabular}
    \label{tab:算法性能比较}
\end{table}

\section{Conclusions and future work}
\label{sect:conclusion}
This study presents a photometric model for the F9 RB that relates the target's attitude to its theoretical brightness, accounting for its physical properties (shape, surface material) and the observation geometry. We introduce the AEAD algorithm, a robust and efficient method for solving the attitude estimation problem. The AEAD algorithm integrates observed light curves with our photometric model, and leverages automatic differentiation to compute the gradient of the resulting loss function, enabling rapid and accurate gradient-based optimization.

The photometric model was also used to generate theoretical light curves for the purpose of validating the AEAD algorithm. By specifying a ground-truth spin axis orientation, we simulated observations for a given location and time period. These simulated light curves served as a testbed for our attitude estimation method. The validation was successful: the estimated attitude parameters demonstrated excellent agreement with the ground-truth values, confirming the accuracy and effectiveness of our approach.

Application of the AEAD algorithm to the observational data from YNAO confirms its significant advantages in parameter optimization. The method exhibits rapid convergence, determining the optimal attitude parameters in approximately 15 seconds while achieving a MSE in the order of $10^{-3}$. Furthermore, a key benefit of our implementation is its deterministic nature. For a given set of initial parameters and a fixed learning rate, the optimization process is fully reproducible and free from stochastic effects. This determinism not only enhances the reliability of the results but also simplifies debugging and validation efforts.

The application of our method to measured light curve data revealed distinct, structured patterns within the solution space of optimal parameters. This finding spurred a deeper investigation into the underlying symmetries of the problem. When the AEAD algorithm was applied to an independent dataset from the MMT observatory, it successfully replicated these characteristic circular distributions, providing strong evidence for the method's generalization capability. Crucially, the ability of AEAD to efficiently process multi-day and multi-site observational data enabled a key scientific discovery: the detection of a dynamic evolution in the target's spin axis orientation. This result not only demonstrates the high sensitivity of our algorithm but also establishes a robust methodological foundation for future research into the long-term rotational dynamics of space objects.

Building upon the work presented here, future research will proceed along several key directions. These include refining the fidelity of the photometric model, exploring more advanced optimization techniques, and enhancing estimation accuracy by incorporating a wider range of observational constraints. Furthermore, a comprehensive comparative analysis will be conducted to benchmark the performance of the AEAD algorithm against other state-of-the-art methods using robust evaluation metrics. Specific avenues for future work include:

1. Error Analysis and Metrics: To perform a rigorous error analysis of the AEAD estimation results and develop more robust evaluation metrics to comprehensively benchmark the algorithm's performance and applicability.

2. Resolving Ambiguities: To incorporate a wider range of geometric constraints—through multi-site simultaneous observations, single-site multi-epoch campaigns, and joint multi-site, multi-epoch observations—to more effectively resolve the solution ambiguities for the spin axis orientation.

3. Model Extension: To extend the AEAD framework to simultaneously invert for key physical properties of the target, such as surface reflectance coefficients and aspect ratio, enabling a more complete physical characterization.

4. Comparative Benchmarking: To conduct a comprehensive performance comparison between the gradient-based AEAD algorithm and alternative optimization methods, such as heuristic algorithms (e.g., PSO, MCMC), to benchmark their relative computational efficiency, convergence properties, and final estimation accuracy.

\begin{acknowledgements}
This work was funded by the funds of the National Natural Science Foundation of China (No. 12373086, No. 12293034, and No. 12003065), the CAS "Light of West China" Program, and the "Yunnan Revitalization Talent Support Program" in Yunnan Province.

Thanks for the open-source Python libraries \texttt{Scipy}, \texttt{Astropy}, \texttt{Jax} and so on.

During the drafting of this article, ChatGPT 3.5 is used to help with translation from Chinese to English.

The data and code involved in this study can be shared with authors who have a reasonable need for it. Light curve data from this research group will be made publicly available in the future.
\end{acknowledgements}


\label{lastpage}


\begin{thebibliography}{99}

\bibitem[Abercrombie et al.(2021)]{Abercrombie+etal+2021} Abercrombie M. D., Calef B., Naderi S., 2021, Advanced Maui Optical and Space Surveillance Technologies Conference (AMOS)

\bibitem[Bradbury et al.(2021)]{Bradbury+etal+2021} Bradbury J., Frostig R., Hawkins P., Johnson M. J., Leary C., Maclaurin D., Necula G., Paszke A., VanderPlas J., Wanderman-Milne S., Zhang Q., 2021, Astrophysics Source Code Library, record ascl:2111.002

\bibitem[Campbell et al.(2022)]{Campbell+etal+2022} Campbell T., Furfaro R., Reddy V., Battle A., Birtwhistle P., Linder T., Tucker S., Pearson N., 2022, The Journal of the Astronautical Sciences, 69(1), 95–119

\bibitem[Clark et al.(2022)]{Clark+etal+2022} Clark R., Fu Y., Dave S., Lee S. K. R., 2022, Advances in Space Research, 70, 3271–3280

\bibitem[Giese(1963)]{Giese+1963} Giese R. H., 1963, SAO Special Report, 127

\bibitem[Huang \& Ji(2016)]{Huang+Ji+2016} Huang P., Ji J., 2016, Acta Astronomica Sinica, 57(5), 568–584

\bibitem[Ji et al.(2023)]{Ji+etal+2023} Ji J., Tan D., Bao C., Huang X., Hu S., Dong Y., and Wang S., 2023, PyMsOfa: A Python Package for the Standards of Fundamental Astronomy (SOFA) Service, Research in Astronomy and Astrophysics, 23:125015 (6pp)

\bibitem[Karpov et al.(2016)]{Karpov+etal+2016} Karpov S., Katkova E., Beskin G., 2016, Revista Mexicana de Astronomiay Astrofisica (Serie de Conferencias) Vol. 48, pp.112-113(2016)

\bibitem[Lin \& Zhao(2018)]{Lin+Zhao+2018} Lin H.-Y., Zhao C.-Y., 2018, Advances in Space Research, 61(1), 182–188

\bibitem[Liu et al.(2019)]{Liu+etal+2019} Liu T., Shen M., Gao P., Zhao Y., 2019, Chinese Journal of Lasers, 46(1), 219–226

\bibitem[Lu \& Zhao(2020)]{Lu+Zhao+2020} Lu Y., Zhao C., 2020, Acta Astronomica Sinica, 61(6), 52–65

\bibitem[Lomb(1976)]{Lomb+1976} Lomb N. R., 1976, Astrophysics and Space Science, 39(2), 447–462

\bibitem[Margossian(2019)]{Margossian+2019} Margossian C. C., 2019, WILEY INTERDISCIPLINARY REVIEWS-DATA MINING AND KNOWLEDGE DISCOVERY, 9(4), e1305

\bibitem[Santoni et al.(2013)]{Santoni+etal+2013} Santoni F., Cordelli E., Piergentili F., 2013, Journal of Spacecraft and Rockets, 50(3), 701–708

\bibitem[Su et al.(2021)]{Su+etal+2021} Su X., Tang R., Li R., Li Y., 2021, Astronomical Research \& Technology, 18(2), 203–212

\bibitem[Scargle(1982)]{Scargle+1982} Scargle J. D., 1982, The Astrophysical Journal, 263(1), 835–853

\bibitem[Tang et al.(2018)]{Tang+etal+2018} Tang R., Li Z., Li Y., Pi X., Su X., Li R., Zhang H., Zhai D., Fu H., 2018, Optics Letters, 43(21), 5488–5491

\bibitem[VanderPlas(2018)]{VanderPlas+2018} VanderPlas J. T., 2018, The Astrophysical Journal Supplement Series, 236, 16

\bibitem[Wang et al.(2008)]{Wang+etal+2008} Wang H., Zhang W., Wang Z., 2008, Acta Optica Sinica, 25(3), 593–598

\bibitem[Williams(1979a)]{Williams+1979a} Williams V., 1979, Planetary and Space Science, 27(6), 885–890

\bibitem[Williams(1979b)]{Williams+1979b} Williams V., 1979, Planetary and Space Science, 27(6), 891–894

\bibitem[Wetterer \& Jah(2009)]{Wetterer+Jah+2009} Wetterer C. J., Jah M., 2009, Journal of Guidance, Control, and Dynamics, 32(5), 1648–1651

\bibitem[Wang et al.(2017)]{Wang+etal+2017} Wang Y., Du X., Fan C., 2017, Chinese Science Bulletin, 62(15), 1578–1590

\bibitem[Yanagisawa \& Kurosaki(2012)]{Yanagisawa+Kurosaki+2012} Yanagisawa T., Kurosaki H., 2012, Advances in Space Research, 50(1), 136–145

\bibitem[Zhao et al.(2020)]{Zhao+etal+2020} Zhao S., Steindorfer M., Kirchner G., Zheng Y., Koidl F., Wang P., Shang W., Zhang J., Li T., 2020, Advances in Space Research, 65(5), 1518–1527

\bibitem[Zhao et al.(2024)]{Zhao+Sun+Yu+2024} Zhao H., Sun R., Yu S., 2024, Deep Neural Network Closed-loop with Raw Data for Optical Resident Space Object Detection, Research in Astronomy and Astrophysics, 24:115009 (9pp). 




















 
\end{thebibliography}
\end{document}